\documentclass[10pt, letterpaper,english,twocolumn,journal]{IEEEtran}
\usepackage[T1]{fontenc}
\usepackage[latin9]{inputenc}
\usepackage{array}
\usepackage{verbatim}
\usepackage{float}
\usepackage{amsthm}
\usepackage{amsmath}
\usepackage{amssymb}
\usepackage{graphicx}
\usepackage{color}
\usepackage{changebar}

\makeatletter

\providecommand{\tabularnewline}{\\}
\floatstyle{ruled}
\newfloat{algorithm}{tbp}{loa}
\providecommand{\algorithmname}{Algorithm}
\floatname{algorithm}{\protect\algorithmname}

\theoremstyle{plain}
\newtheorem{thm}{\protect\theoremname}
\theoremstyle{remark}
\newtheorem{rem}[thm]{\protect\remarkname}

\IEEEoverridecommandlockouts
\usepackage{ifpdf}\usepackage{cite}
\ifCLASSOPTIONcompsoc
\else
\fi

\usepackage{babel}
\providecommand{\remarkname}{Remark}
\providecommand{\theoremname}{Theorem}

\makeatother

\usepackage{babel}
\providecommand{\remarkname}{Remark}
\providecommand{\theoremname}{Theorem}

\begin{document}

\title{Secrecy Transmit Beamforming for Heterogeneous Networks}

\author{Tiejun Lv, \emph{Senior Member, IEEE}, Hui Gao, \emph{Member, IEEE},
and Shaoshi Yang, \emph{Member, IEEE} \emph{ }%
\thanks{Manuscript received July 24, 2014; revised December 17, 2014; accepted February 19, 2015. 

This work is financially supported by the National Natural Science
Foundation of China (NSFC) (Grant No. 61271188 and Grant No. 61401041) and the Fundamental
Research Funds for the Central Universities (Grant No. 2014RC0106).%
} %
\thanks{T. Lv, and H. Gao are with the School of Information and Communication
Engineering, Beijing University of Posts and Telecommunications, Beijing,
China 100876 (e-mail: \{lvtiejun, huigao\}@bupt.edu.cn).%
} %
\thanks{S. Yang (Correspondence Author) is with the School of Electronics and Computer Science, University
of Southampton, Southampton, SO17 1BJ, U.K. (e-mail: sy7g09@ecs.soton.ac.uk).%
}}
\maketitle
\begin{abstract}
In this paper, we pioneer the study of physical-layer security in heterogeneous networks (HetNets). We investigate secure communications in a two-tier downlink HetNet, which comprises one macrocell and several femtocells. Each cell has multiple users and an eavesdropper attempts to wiretap the intended macrocell user. Firstly, we consider an orthogonal spectrum allocation strategy to eliminate co-channel interference, and propose the secrecy transmit beamforming only operating in the macrocell (STB-OM) as a partial solution for secure communication in HetNet. Next, we consider a secrecy-oriented non-orthogonal spectrum allocation strategy and propose two cooperative STBs which rely on the collaboration amongst the macrocell base station (MBS) and the adjacent femtocell base stations (FBSs). Our first cooperative STB is the STB sequentially operating in the macrocell and femtocells (STB-SMF), where the cooperative FBSs individually design their STB matrices and then feed their performance metrics to the MBS for guiding the STB in the macrocell. Aiming to improve the performance of STB-SMF, we further propose the STB jointly designed in the macrocell and femtocells (STB-JMF), where all cooperative FBSs feed channel state information to the MBS for designing the joint STB. Unlike conventional STBs conceived for broadcasting or interference channels, the three proposed STB schemes all entail  relatively sophisticated optimizations due to QoS constraints of the legitimate users. In order to efficiently use these STB schemes, the original optimization problems are reformulated and convex optimization techniques, such as second-order cone programming and semidefinite programming, are invoked to obtain the optimal solutions. Numerical results demonstrate that the proposed STB schemes are highly effective in improving the secrecy rate performance of HetNet.

\end{abstract}


\section{Introduction}

\global\long\def\figurename{Fig.}
With the rapid proliferation of smart phones, tablets and machine-to-machine communications, there is an ever-increasing
demand for seamless wireless coverage, extremely high mobile data rate and reliable secrecy performance
\cite{_cisco_????}. Aiming to effectively enhance the spectral and/or energy efficiency
of wireless networks, it has been suggested that the deployment density of various network-nodes should increase \cite{boccardi_five_2014,chandrasekhar_femtocell_2008}, and smart access-point activation and resource management are essential to utilize the dense network-nodes \cite{liao_base_2014}.
Therefore, heterogeneous network (HetNet) \cite{damnjanovic_survey_2011,ghosh_heterogeneous_2012} has
been attracting great research interests, and it is regarded as one of the most promising techniques
for providing higher network capacity and wider coverage. HetNet is supported by various types of base stations with different transmit power budgets.
Macrocell base stations (MBSs) provide public access and wide area
coverage up to a few kilometers to all the marcocell users (MUs). Small
cell base stations, such as femtocell base stations (FBSs), are typically
overlaid on the existing macrocells and installed in the building
or indoor environment close to femtocell users (FUs). It is noted that the network architecture of HetNet becomes more open and diverse compared to the conventional single-tier cellular networks, which
makes the information exchange more susceptible to eavesdropping.
Recently, physical-layer security (PLS) has been proposed as a family of viable techniques to
secure wireless communications \cite{csiszar1978broadcast,barros_secrecy_2006}, and it also has the potential to tackle the security problem encountered in HetNet.

Traditionally, the security problem in wireless networks was mainly studied at higher layers using key-based cryptographic methods \cite{shannon1949communication}. This conventional wisdom relies on the assumption that the eavesdropper is not computationally powerful to break the secret key. However, as the computational capability of wireless devices develops rapidly, perfect security can be hardly guaranteed with the key-based solutions. It is reasonable to argue that security measures should be invoked at all layers where they can be implemented in a cost-effective manner. Notably, the physical layer has remained almost ignored for security in the past.
The basic idea of PLS is to exploit
the randomness of wireless channel for secure message transmission \cite{shiu_physical_2011}. The authors in \cite{barros_secrecy_2006,liang2008secure,gopala2008secrecy} investigated the PLS
in single-antenna fading channels. Since then, the secure communication
in multi-antenna channels has been extensively studied \cite{hero2003secure,bustin2009mmse,mukherjee2009fixed,liu2009note,khisti2010secure,oggier2011secrecy,cumanan_secrecy_2014} and in particular, practical and robust multi-antenna secrecy beamforming schemes were investigated in \cite{cumanan_secrecy_2014} very recently.
Secure broadcasting with more than two receivers were considered in
\cite{khisti2008secure,ekrem2011secrecy,liu2010vector}. Other related works in the contexts of the multiple-access channel with confidential messages \cite{liang2008multiple},
the multiple-access channel wiretap channel \cite{tekin2008gaussian}, and the cognitive
multiple-access channel with confidential messages \cite{liu2011fading} have also been reported. Recently, there are growing research interests in the secrecy communication
over interference channels, where the interference may be potentially exploited for improving the secrecy performance.
More specifically, in \cite{liu2008discrete}, the authors studied the inner and outer bounds for the secrecy
capacity regions of a two-user interference network where the receivers
are potentially eavesdroppers. The secrecy rate of a two-user interference
network with an outer eavesdropper was investigated in \cite{agrawal2009secrecy}.
The MIMO Gaussian interference channel with confidential messages
was studied in \cite{fakoorian2011mimo} using game-theoretic approach.
The authors of \cite{ni2013robust} addressed the problem of minimizing the transmit power
with imperfect channel state information (CSI) in a $K$-user multiple-input single-output interference network,
and the so-called "S-procedure" was applied.
It should be noted that the existing works on PLS mainly
focus on traditional network architectures, and the research on PLS for HetNet is still largely missing.

Because of the densely overlaid network architecture, there is ubiquitous interference of various types existing in HetNet \cite{lopez-perez_enhanced_2011}. From the
viewpoint of PLS, deliberately introducing interference
can be beneficial for the secrecy rate performance of the system \cite{shiu_physical_2011}.
Inspired by this insight, the interference should be utilized, rather
than avoided, to improve the secrecy rate in HetNet using techniques such as proper spectrum allocation and cooperative beamforming. To elaborate a little further, motivated by the inter-cell
interference coordination techniques \cite{niyato_competitive_2008,de_lima_coordination_2012,lopez-perez_distributed_2013,lopez-perez_power_2014,pantisano_improving_2014},
spectrum allocation can be rearranged dynamically in conjunction with various levels
of cooperation between the network nodes, such as the MBS and FBSs, to generate the desired co-channel
interference (CCI) to the eavesdropper. As a result, judicious cooperative beamforming
schemes may be designed to cope with CCI for the sake of improving the secrecy rate performance in HetNet.

In this paper, a two-tier downlink HetNet system is considered, where MBS and FBSs serve the corresponding legitimate MUs and FUs, and a MU acts maliciously as an eavesdropper to wiretap another legitimate MU. For the considered scenario, we propose three secrecy transmit beamforming (STB) schemes in conjunction with two spectrum allocation schemes for secure communications in the HetNet, assuming different degrees of cooperation among the network nodes. As an initial study, we first consider the conventional orthogonal spectrum allocation (OSA) strategy \cite{heath_modeling_2013}, where orthogonal spectrum resources are allocated to the MBS and the adjacent FBSs to eliminate the cross-tier interference and the interference among adjacent femtocells. Employing OSA, the considered scenario can be simplified as the secure communication in a broadcasting channel, and we consider the STB only performed in macrocell (STB-OM) as a partial solution for the secure communication in HetNet. It is noted that STB-OM aims to maximize the secrecy rate of the intended MU subject to the Quality of Service (QoS) constraints of the other legitimate MUs; no cooperation between the MBS and FBSs is necessary. Inspired by the fact that friendly interference can help secure communication \cite{shiu_physical_2011}, we deliberately introduce interference in the HetNet with the secrecy-oriented non-orthogonal spectrum allocation (SONOSA) strategy, which dynamically changes the local pattern of the underlay OSA. Specifically, some cooperative FBSs adjacent to the eavesdropper are assigned with the same frequency resource as MBS, while the OSA strategy still applies to the non-cooperative FBSs. Consequently, CCI is deliberately introduced around the eavesdropper to enable more effective cooperative STB. Specifically, two STB schemes are proposed in conjunction with SONOSA, where MBS performs STB in collaboration with its cooperative co-channel FBSs. Firstly, a STB scheme is proposed to improve STB-OM with a little cross-tier cooperation, which is sequentially performed in macrocell and femtocells (STB-SMF). In this scheme, each cooperative FBS designs its STB to altruistically maximize the generated interference to the eavesdropper while serving its own FUs. Then each cooperative FBS calculates a performance metric and feeds it back to the MBS to facilitate the STB in macrocell, and the MBS can still use the STB-OM with minor modification. In
order to improve the overall performance of STB-SMF, another STB is proposed with the limited cross-tier cooperation, which is jointly performed in macrocell and femtocells (STB-JMF). In this scheme, each cooperative FBS feeds its CSI back to the MBS for the joint STB, which aims to guarantee the QoS requirements of both MUs and FUs while enhancing the secrecy rate of the intended MU.

For the sake of clarity, the main contributions of this paper are summarized as follows:
\begin{enumerate}
\item Upon adopting the OSA strategy, an STB-OM scheme is proposed
to secure the broadcast channel of the macrocell, where the confidential message and the common messages are simultaneously transmitted. In particular, an iterative algorithm is proposed to maximize the secrecy rate while satisfying the QoS of the common messages. In order to handle the complicated optimization problem, we transform the original nonconvex problem
into a second-order cone program (SOCP) with the aid of a first-order Taylor
approximation. The approximation can be improved with each iteration.
Therefore, a local optimum of the original optimization problem can be obtained
in several iterations. It is noted that the proposed STB-OM is different from the conventional STB schemes in the broadcasting channel \cite{hanif2014linear, geraci2014secrecy}, where only the confidential message transmission is considered. Furthermore, the STB-OM is applicable to the general single cell scenarios and does not need cooperation between the MBS and FBSs, which also initiates our cooperative STB exploiting CCI.

\item Based on the SONOSA strategy, a STB-SMF scheme is proposed to improve the secrecy rate of STB-OM while guaranteeing the QoS requirements of legitimate MUs. Different from the traditional jamming sources \cite{zheng2011optimal}, the cooperative FBSs still serve their FUs while helping the MBS. In order to enhance the secrecy rate performance of the intended MU, the cooperative FBSs selflessly increase the interference power towards the eavesdropper without considering the QoS of their FUs, and the optimization problem is efficiently solved by SOCP. It is worth pointing out that each cooperative FBS designs its STB with local CSI, and only a scalar is fed back to the MBS for its STB in macrocell, where the MBS can still adopt STB-OM. Therefore, the STB-SMF scheme imposes very little overhead for the cross-tier cooperation, and is compatible with STB-OM.

\item Employing SONOSA strategy, an STB-JMF scheme is proposed to strike a better balance between the secret-rate performance of the intended MUs and the QoS requirements of the legitimate MUs and FUs. It is noted that such complicated optimization is nonconvex, and we opt for reformulating it into a tractable two-stage problem. Specifically, we first fix the signal-to-interference-plus-noise ratio (SINR) of the eavesdropper to formulate the inner stage problem, which can be further transformed into a tractable semidefinite program (SDP) by applying the semidefinite relaxation technique. Then, we perform a one-dimensional search to solve the outer stage problem, which finds the most suitable SINR of the eavesdropper to optimize the original objective. It is noted that the MBS only needs to collect the CSI from its cooperative FBSs for this joint STB, and such cross-tier cooperation imposes acceptable overhead for the backhaul. Furthermore, unlike the altruistic STB of the cooperative FBSs in STB-SMF, the STB-JMF shceme guarantees the QoS of the FUs in the cooperative FBSs.

\end{enumerate}

The rest of this paper is organized as follows. In Section II, the
downlink HetNet system model and the corresponding spectrum allocation strategies
are presented, and we derive SINR expressions of various network nodes for the
following beamforming design. Based on the two spectrum allocation
strategies, three STB schemes are proposed in Section III, where the related optimization problems are formulated and solved. In Section IV, simulation results show the effectiveness of the proposed algorithms. Finally, conclusions and future directions are provided in Section V.

$Notations$: Bold upper and lower case letters denote matrices and
vectors, respectively. Transpose and conjugate transpose are denoted
by $\left(\cdot\right)^{T}$ and $\left(\cdot\right)^{H}$, respectively, while $\mathrm{E}\left\{ \cdot\right\} $
represents expectation. $\mathrm{Tr}\left(\cdot\right)$ is the trace
operator, $\|\cdot\|$ represents the Euclidean norm, $|\cdot|$ denotes
the mode of a complex number, and $\mathrm{rank}\left(\cdot\right)$
stands for the rank of a matrix. $\mathbf{X}\succeq0$ indicates that $\mathbf{X}$
is Hermitian positive semidefinite. $\mathbb{C}$ represents the field
of complex numbers. $\mathrm{Re}\left(\cdot\right)$ and $\mathrm{Im}\left(\cdot\right)$
denote the real part and the imaginary part of a complex number, respectively.
Additionally, the integer set $\left\{ 1,2,\ldots,K\right\} $ is abbreviated as
$\left[1,K\right]$. $\mathcal{CN}\left(\mu,\sigma^{2}\right)$ denotes
a complex Gaussian variable with mean $\mu$ and variance $\sigma^{2}$.

\section{System Model}
\begin{figure}[t]
\begin{centering}
\includegraphics[width=8cm]{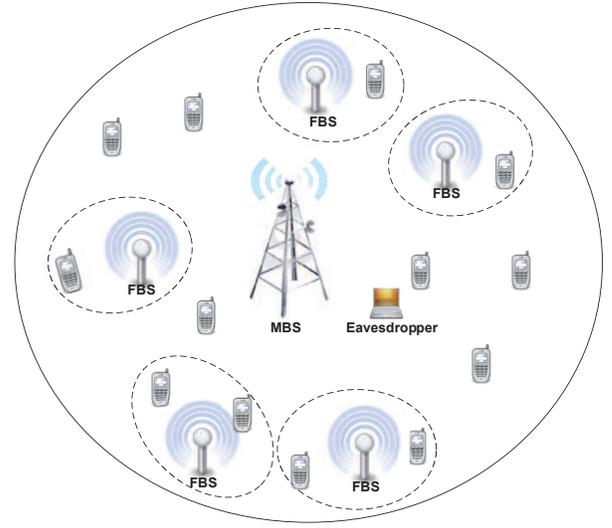}
\par\end{centering}
\caption{Downlink HetNet system model, comprising a macrocell and $N$ femtocells.
The macrocell consists of a MBS, $M$ legitimate MUs and an eavesdropper.
Each FBS serves $K$ FUs. }
\label{system_totol}
\end{figure}

We consider a downlink HetNet as shown in Fig. 1. There is one $N_{M}$-antenna
MBS at the center, and $M$ single-antenna MUs are randomly distributed
throughout the macrocell coverage area, where we have $N_{M}>M$, while the single-antenna eavesdropper
intends to wiretap the confidential message transmitted to a legitimate
MU. Additionally, $F$ $N_{F}$-antenna FBSs are spatially distributed
according to a homogeneous Poisson point process and each FBS
aims to serve $K$ single-antenna FUs, where we have $N_{F}>K$. To simplify the analysis,
we assume that the transmit power of each FBS is fixed and equal, denoted
by $P_{F}$. Similar to that of the FBSs, the transmit power of the MBS is assumed
to be $P_{M}$.  Other relevant
variables are defined in Table 1.

Two spectrum allocation strategies are developed in this paper, i.e., the OSA strategy and the SONOSA strategy. Serving as the underlay spectrum allocation strategy, OSA allocates orthogonal spectrum resources to the MBS and the adjacent FBSs to eliminate the cross-tier interference as well as the interference among adjacent femtocells \cite{heath_modeling_2013}. Building upon OSA, SONOSA dynamically changes the local pattern of OSA to enable advanced cooperative STB schemes. Specifically, the SONOSA strategy assigns the same frequency resource occupied by the MBS to the FBSs which are adjacent to the eavesdropper, as shown in Fig. 2. Employing SONOSA, the co-channel FBSs can work in collaboration with the MBS to generate interference to the eavesdropper while serving their own FUs.

\vspace{-0.2cm}
\begin{table}[tbp]
\centering{}\caption{List of The Major Variables}
\begin{tabular}{|>{\centering}p{2cm}|>{\raggedright}p{5.5cm}|}
\hline
Variable  & Definition\tabularnewline
\hline
\hline
$\mathrm{MU}_{m}$, $\mathrm{\mathrm{FB}S}_{n}$ $\mathrm{FU}_{nk}$  & the $m$-th MU, the $n$-th cooperative FBS of MBS and the $k$-th FU of the $n$-th cooperative FBS\tabularnewline
$s_{m}$  & the message signal intended for the $\mathrm{MU}_{m}$, $m\in[1,M]$
, satisfying $\mathrm{E}\{\left\Vert s_{m} \right\Vert\}=1 $\tabularnewline
$s_{nk}$  & the message signal intended for $\mathrm{FU}_{nk}$, $n\in\left[1,N\right]$,
$k\in\left[1,K\right]$, satisfying $\mathrm{E}\{\left\Vert s_{nk}\right\Vert\}=1 $\tabularnewline
$\mathbf{h}_{m}$  & the channel vector from the MBS to $\mathrm{MU}_{m}$\tabularnewline
$\mathbf{h}_{n,m}$  & the channel vector from $\mathrm{\mathrm{FB}S}_{n}$ to $\mathrm{MU}_{m}$\tabularnewline
$\mathbf{h}_{E}$  & the channel vector from the MBS to the eavesdropper\tabularnewline
$\mathbf{h}_{n,E}$  & the channel vector from $\mathrm{\mathrm{FB}S}_{n}$ to the eavesdropper\tabularnewline
$\mathbf{h}_{nk}$  & the channel vector from the MBS to $\mathrm{FU}_{nk}$\tabularnewline
$\mathbf{h}_{p,nk}$  & the channel vector from $\mathrm{\mathrm{FB}S}_{p}$ to $\mathrm{FU}_{nk}$, $p\in\left[1,N\right]$\tabularnewline
$\mathbf{w}_{m}$  & the precoding vector for $\mathrm{MU}_{m}$\tabularnewline
$\mathbf{w}_{nk}$  & the precoding vector for $\mathrm{FU}_{nk}$\tabularnewline
\hline
\end{tabular}
\end{table}

\begin{figure}[t]
\centering{}\includegraphics[scale=0.45]{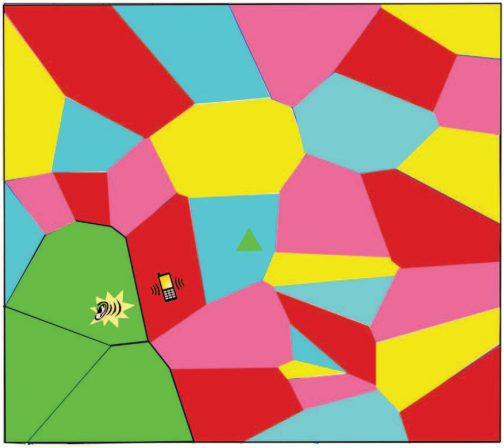}\caption{An example for OSA and SONOSA, where
the polygons with the same color denote the femtocells with the same
frequency resource, the apple-green triangle is MBS, the cell phone is the
eavesdropped MU and the ear denotes the eavesdropper. It is noted that OSA is the underlay strategy, and SONOSA dynamically changes the local pattern of OSA. Three apple-green femtocells are assigned with the same frequency resource as the macrocell to generate CCI against the eavesdropper.}
\end{figure}

Considering SONOSA, we assume that there are $N$ ($1\leq N<F $) cooperative FBSs employing CCI to improve the secrecy rate of the eavesdropped MU. Let us denote the $n$-th cooperative FBS of the MBS as $\mathrm{FBS}_{n}$ and the $m$-th MU as $\mathrm{MU}_{m}$, then the received signal at $\mathrm{MU}_{m}$ is given by
\begin{equation}\begin{split}
y_{m}= & \mathbb{\mathbf{h}}_{m}\mathbf{w}_{m}s_{m}+\sum_{q=1,q\neq m}^{M}\mathbb{\mathbf{h}}_{m}\mathbf{w}_{q}s_{q}\\&+\sum_{n=1}^{N}\sum_{k=1}^{K}\mathbb{\mathbf{h}}_{n,m}\mathbf{w}_{nk}s_{nk}+\mathbf{\mathit{\mathbf{\mathit{n}}}}_{m},\,\, m\in\left[1,M\right],
\end{split}\end{equation}
where $\mathbb{\mathbf{h}}_{m}\in\mathbb{C}^{1\times N_{M}}$ and $\mathbb{\mathbf{h}}_{n,m}\in\mathbb{C}^{1\times N_{F}}$
denote the channel vector from the MBS to $\mathrm{MU}_{m}$ and the
channel vector from the $\mathrm{FBS}_{n}$ to $\mathrm{MU}_{m}$, respectively.\footnote{In this paper, we just assume perfect CSI as our first
step to get fundamental insights into the physical layer security problem
in HetNet. We will consider a more practical model including imperfect
CSI in our future work.}
$\mathbf{w}_{m}\in\mathbb{C}^{N_{M}\times1}$ and $s_{m}$ represent
the precoding vector and the message symbol intended for $\mathrm{MU}_{m}$, respectively. Similarly,
$\mathbf{w}_{nk}\in\mathbb{C}^{N_{F}\times1}$ and $s_{nk}$ are the precoding vector and the message signal intended for the $k$-th FU of the $n$-th cooperative FBS, denoted as $\mathrm{FU}_{nk}$. $\mathbf{\mathit{\mathbf{\mathit{n}}}}_{m}$ is the
Gaussian noise following independent and identically distributed (i.i.d.)
$\mathcal{CN}\left(0,\sigma_{M}^{2}\right)$ at $\mathrm{MU}_{m}$.
The precoding vectors for the MUs satisfy the power constraint of $\sum_{m=1}^{M}\left\Vert \mathbf{w}_{m}\right\Vert ^{2}=P_{M}$
and the precoding vectors for the FUs in a femtocell satisfy the power constraint of $\sum_{k=1}^{K}\left\Vert \mathbf{w}_{nk}\right\Vert ^{2}=P_{F}$.
Without loss of generality, we assume that there is an eavesdropper
wiretapping $\mathrm{MU}_{1}$\footnote{Similar to \cite{kalantari2014joint}, we focus on the beamforming design in this paper. The identification problem -- how to identify which MU has been intercepted by the eavesdropper -- is challenging and important, which will be investigated in detail in our future work. In principle, this problem can be solved by authentication: the MBS may broadcast some test messages to each MU in a round-robin manner, and ask each of them to feed back what has been listened. For the wiretapped MU, the MBS will receive feedback from two sources.}, therefore the received signal at the eavesdropper
is given by
\begin{equation}\begin{split}
y_{E}= & \mathbb{\mathrm{\mathbf{h}}}_{E}\mathbf{w}_{1}s_{1}+\sum_{m=2}^{M}\mathbb{\mathbf{h}}_{E}\mathbf{w}_{m}s_{m}\\&+\sum_{n=1}^{N}\sum_{k=1}^{K}\mathbb{\mathbf{h}}_{n,E}\mathbf{w}_{nk}s_{nk}+\mathbf{\mathit{n}}_{E},
\end{split}\end{equation}
where $\mathbb{\mathbf{h}}_{E}\in\mathbb{C}^{1\times N_{M}}$ denotes
the channel vector from the MBS to the eavesdropper, $\mathbb{\mathbf{h}}_{n,E}\in\mathbb{C}^{1\times N_{F}}$
is the channel vector from $\mathrm{FBS}_{n}$ to the eavesdropper, and $\mathbf{\mathit{n}}_{E}$
is the Gaussian noise obeying i.i.d. $\mathcal{CN}\left(0,\sigma_{E}^{2}\right)$
at the eavesdropper. Furthermore, the received signal at $\mathrm{FU}_{nk}$ can be formulated as
\begin{equation}\begin{split}
y_{nk}&=\mathbb{\mathbf{h}}_{n,nk}\mathbf{w}_{nk}s_{nk}+\sum_{t=1,t\neq k}^{K}\mathbb{\mathbf{h}}_{n,nk}\mathbf{w}_{nt}s_{nt}\\&+\sum_{p=1,p\neq n}^{N}\sum_{t=1}^{K}\mathbb{\mathbf{h}}_{p,nk}\mathbf{w}_{pt}s_{pt}+\sum_{m=1}^{M}\mathbb{\mathbf{h}}_{nk}\mathbf{w}_{m}s_{m}+\mathbf{\mathit{n}}_{nk},
\end{split}\end{equation}
where $\mathbb{\mathbf{h}}_{n,nk}\in\mathbb{C}^{1\times N_{F}}$ is
the channel vector from $\mathrm{FBS}_{n}$ to $\mathrm{FU}_{nk}$, $\mathbf{h}_{nk}\in\mathbb{C}^{1\times N_{M}}$ is the channel vector
form $\mathrm{MBS}_{n}$ to $\mathrm{FU}_{k}$,
and the Gaussian noise $\mathbf{\mathit{n}}_{nk}$ at $\mathrm{FU}_{nk}$
follows i.i.d. $\mathcal{CN}\left(0,\sigma_{F}^{2}\right)$.

To facilitate the analysis of secrecy rate performance, we define the SINR
of $\mathrm{MU}_{m}$ as

\begin{equation}
\mathrm{SINR}_{m}=\frac{\left|\mathbb{\mathbf{h}}_{m}\mathbf{w}_{m}\right|{}^{2}}{A_{m}},m\in\left[1,M\right],\label{eq:SINRq}
\end{equation}
where
\[
A_{m}=\sum_{q=1,q\neq m}^{M}\left|\mathbb{\mathbf{h}}_{m}\mathbf{w}_{q}\right|^{2}+\sum_{n=1}^{N}\sum_{k=1}^{K}\left|\mathbb{\mathbf{h}}_{n,m}\mathbf{w}_{nk}\right|^{2}+\sigma_{M}^{2}.
\]
Similarly, the SINR of the eavesdropper is represented as
\begin{equation}
\mathrm{SINR}_{E}=\frac{\left|\mathbb{\mathbf{h}}_{E}\mathbf{w}_{1}\right|^{2}}{B_{E}},\label{eq:SINRe}
\end{equation}
where
\[
B_{E}=\sum_{m=2}^{M}\left|\mathbb{\mathbf{h}}_{E}\mathbf{w}_{m}\right|^{2}+\underbrace{\sum_{n=1}^{N}\sum_{k=1}^{K}\left|\mathbb{\mathbf{h}}_{n,E}\mathbf{w}_{nk}\right|^{2}}_{\mathrm{IFT}_{sum}}+\sigma_{E}^{2},
\]
and $\mathrm{IFT}_{sum}$ is the sum of the additional interference temperature from all the cooperative FBSs to the eavesdropper. Finally, the SINR of the $\mathrm{FU}_{nk}$ is written as
\begin{equation}
\mathrm{SINR}_{nk}=\frac{\left|\mathbb{\mathbf{h}}_{n,nk}\mathbf{w}_{nk}\right|^{2}}{C_{nk}},n\in\left[1,N\right],k\in\left[1,K\right],
\end{equation}
where
\begin{align*}
C_{nk}= & \sum_{t=1,t\neq k}^{K}\left|\mathbb{\mathbf{h}}_{n,nk}\mathbf{w}_{nt}\right|^{2}+\sum_{p=1,p\neq n}^{N}\sum_{t=1}^{K}\left|\mathbb{\mathbf{h}}_{p,nk}\mathbf{w}_{pt}\right|^{2}\\&+\sum_{m=1}^{M}\left|\mathbb{\mathbf{h}}_{nk}\mathbf{w}_{m}\right|^{2}+\sigma_{F}^{2}.
\end{align*}

After the above preliminary derivations, we will propose three STB schemes in the next section to maximize the secrecy rate of the intended $\mathrm{MU}_{1}$.

\section{Security Transmit Beamforming Schemes for HetNet}

In this section, we propose three secrecy communication schemes to
improve the secrecy rate of HetNet by jointly considering the spectrum
allocation strategy and the transmit beamforming. Firstly, based on the
OSA strategy, we propose the STB-OM scheme as a partial solution for
secure communication in HetNet. Then, we employ the SONOSA strategy
to propose the STB-SMF scheme, which improves STB-OM with secrecy-oriented
CCI generated by the altruistic cooperative FBSs. At last, aiming at
a balanced system performance, we propose the STB-JMF scheme, which
optimizes the security-rate of the intended MU while satisfying
the QoS constraints of both legitimate MUs and FUs. In the remaining
part of this section, we will discuss each STB scheme in detail. Note that in all the three proposed schemes, we assume that the local CSI of the
MUs and of the eavesdropper is available at the MBS\footnote{Similar to \cite{cumanan_secrecy_2014,mukherjee2009fixed,bustin2009mmse}, and\cite{hanif2014linear}, we assume that the CSI of the eavesdropper is also available at the MBS in order to make the STB more
tractable.}, and each FBS knows the local CSI of its own femtocell.

\subsection{Secrecy Transmit Beamforming Only Performed in Macrocell (STB-OM)\label{sub:Secrecy-Transmit-Beamforming-1}}

Let us first discuss the STB-OM scheme that relies on the OSA strategy. Because different frequency resources are allocated
to the MBS and FBSs, no CCI exists in this scheme.
As shown in Fig. 3, the MBS serves multiple legitimate MUs, and we
assume that $\text{MU}_{1}$ is wiretapped by the eavesdropper. Our
goal is to maximize the secrecy rate of $\mathrm{MU}_{1}$ while guaranteeing the QoS, i.e., SINR, requirements of other legitimate
MUs. The proposed STB-OM scheme is aiming at the secrecy rate maximization.
Unfortunately, this optimization problem, formulated as (7a) to (7c),
is non-convex and hence prohibitive computational complexity may be
incurred for finding its optimum solution. As a compromise, we employ
a low-complexity iterative algorithm \cite{Vishwanathan05kernelmethods}
which conservatively approximates the original problem as several
tractable SOCP subproblems. In the remaining part of this subsection,
the details of the problem formulation and solution
are provided.

\begin{figure}[t]
\begin{centering}
\includegraphics[width=8cm]{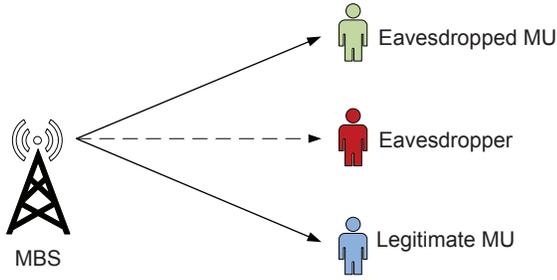}\caption{An example channel model for the STB-OM scheme, consisting of one
MBS, $M=2$ MUs and one eavesdropper wiretaps the first MU. As the
OSA is employed, the system is equivalent
to a broadcast system. The solid lines indicate useful data streams,
and the dash line indicates the interference stream. }

\par\end{centering}

\label{system_broadcast-1}
\end{figure}

Initially, the secrecy rate optimization problem of STB-OM can be formulated
as\begin{subequations}\label{SRM1}
\begin{eqnarray}
\max_{\{\mathbf{w}_{m}\}_{m=1}^{M}} & \log(1+\mathrm{SINR}_{1})-\log(1+\mathrm{SINR}_{E})\\
\mathrm{s.t.}\;\; & \mathrm{SINR}_{m}\ge\gamma_{m},m\in\left[2,M\right],\label{SRM1a}\\
 & \sum_{m=1}^{M}\left\Vert \mathbf{w}_{m}\right\Vert ^{2}\le P_{M},\label{SRM1b}
\end{eqnarray}
\end{subequations}where the received SINRs from (\ref{eq:SINRq}) and (\ref{eq:SINRe}),
that are associated with $\mathrm{MU}_{m}$ and the eavesdropper, can be simplified
as
\begin{equation}
\mathrm{SINR}_{m}=\frac{\left|\mathbb{\mathbf{h}}_{m}\mathbf{w}_{m}\right|^{2}}{\sigma_{M}^{2}+\sum_{q=1,q\neq m}^{M}\left|\mathbb{\mathbf{h}}_{m}\mathbf{w}_{q}\right|^{2}},\label{SINRbq}
\end{equation}
and
\begin{equation}
\mathrm{SINR}_{E}=\frac{\left|\mathbb{\mathbf{h}}_{E}\mathbf{w}_{1}\right|^{2}}{\sigma_{E}^{2}+\sum_{m=2}^{M}\left|\mathbb{\mathbf{h}}_{E}\mathbf{w}_{m}\right|^{2}+\mathrm{IFT}_{sum}},\label{SINRbe}
\end{equation}
respectively, when the OSA strategy is adopted.
Note that the $\mathrm{IFT}_{sum}$ in this STB-OM is zero, because OSA does not introduce CCI. The constraint (\ref{SRM1a})
is the QoS requirements of the other legitimate MUs, i.e., $\mathrm{MU}_{m},m\in\left[2,M\right]$,
and (\ref{SRM1b}) represents the total transmit power constraint
at the MBS. To make the above problem more tractable, we introduce
a pair of slack variables $t_{1}$ and $t_{2}$. 
Then, the original problem can be equivalently reformulated as \begin{subequations}\label{SRM2}
\begin{align}
\max_{\{\mathbf{w}_{m}\}_{m=1}^{M},t_{1},t_{2}} & \quad\log(t_{1})+\log(t_{2})\\
\mathrm{s.t.}\quad\  & \quad1+\mathrm{SINR}_{1}\ge t_{1},\label{SRM2a}\\
 & \quad1+\mathrm{SINR}_{E}\le1/t_{2},\label{SRM2b}\\
 & \quad\mathrm{SINR}_{m}\ge\gamma_{m},m\in\left[2,M\right],\\
 & \quad\sum_{m=1}^{M}\left\Vert \mathbf{w}_{m}\right\Vert ^{2}\le P_{M}.
\end{align}
\end{subequations} Without loss of generality, in the following,
we assume $\sigma_{M}^{2}=\sigma_{E}^{2}=\sigma_{F}^{2}=\sigma^{2}=1.$
Substituting (\ref{SINRbq}) and (\ref{SINRbe}) into the above problem,
(\ref{SRM2}) can be transformed as follows \begin{subequations}\label{SRM3}
\begin{align}
\max_{\{\mathbf{w}_{m}\}_{m=1}^{M},t_{1},t_{2}} & \quad t_{1}t_{2}\\
\mathrm{s.t.}\quad\  & \quad1+\sum_{m=2}^{M}\mathbf{w}_{m}^{H}\mathbf{H}_{1}\mathbf{w}_{m}\le\frac{\mathbf{w}_{1}^{H}\mathbf{H}_{1}\mathbf{w}_{1}}{t_{1}-1},\label{SRM3b}\\
 & \quad1+\sum_{m=1}^{M}\mathbf{w}_{m}^{H}\mathbf{H}_{E}\mathbf{w}_{m}\nonumber\\&\quad\le\frac{1+\sum_{m=2}^{M}\mathbf{w}_{m}^{H}\mathbf{H}_{E}\mathbf{w}_{m}}{t_{2}},\label{SRM3c}\\
 & \quad\gamma_{m}\left(1+\sum_{q=1,q\ne m}^{M}\mathbf{w}_{q}^{H}\mathbf{H}_{m}\mathbf{w}_{q}\right)\nonumber\\&\quad\le\mathbf{w}_{m}^{H}\mathbf{H}_{m}\mathbf{w}_{m},m\in\left[2,M\right],\\
 & \quad\sum_{m=1}^{M}\left\Vert \mathbf{w}_{m}\right\Vert ^{2}\le P_{M},
\end{align}
\end{subequations}where we introduce the matrices $\mathbf{H}_{1}=\mathbf{h}_{1}\mathbf{h}_{1}^{H}$,
$\mathbf{H}_{m}=\mathbf{h}_{m}\mathbf{h}_{m}^{H}$, $\mathbf{H}_{E}=\mathbf{h}_{E}\mathbf{h}_{E}^{H}$.
It is noted the logarithm function is removed because of its monotonically
increasing property. As we can see, the right-hand side of the constraints
(\ref{SRM3b}) and (\ref{SRM3c}) are both quadratic-over-linear functions
which are convex \cite{boyd2004convex}. However, both constraints remain non-convex because they are expressed as the form of $f_1(x)\le f_2(x)$, namely $f_1(x)-f_2(x)\le 0$, where $f_1(x)$ and $f_2(x)$ are both convex functions. More specifically, it is well known that the sum of convex functions is convex. Then, $f_1(x)-f_2(x)$ is non-convex. Following the idea of the constrained convex procedure\cite{Vishwanathan05kernelmethods}, we replace these quadratic-over-linear functions with their corresponding first-order expansions, and then the problem can be transformed into a convex one. Specifically, we define
\begin{equation}
F_{\mathbf{A},a}(\mathbf{w},x)=\frac{\mathbf{w}^{H}\mathbf{A}\mathbf{w}}{x-a},\label{eq:FA}
\end{equation}
where $x\ge a,\mathbf{A}\succeq0$. The first-order Taylor
expansion of (\ref{eq:FA}) at a certain point ($\tilde{\mathbf{w}},\tilde{x}$)
is given by
\begin{equation}
Q_{\mathbf{A},a}(\mathbf{w},x,\tilde{\mathbf{w}},\tilde{x})=\frac{2\mathrm{Re}\{\tilde{\mathbf{w}}^{H}\mathbf{A}\mathbf{w}\}}{\tilde{x}-a}-\frac{\tilde{\mathbf{w}}^{H}\mathbf{A}\tilde{\mathbf{w}}}{(\tilde{x}-a)^{2}}(x-a).
\end{equation}
Furthermore, it is noted that $\max\, t_{1}t_{2}$ can transformed
into a SOCP representation, e.g., $\max\, t_{0}$ with an additional
constraint $\left\Vert \left[2t_{0}\,(t_{1}-t_{2})\right]\right\Vert \le t_{1}+t_{2}$.
This transformation is based on the fact that the constraint $t_{1}t_{2}\ge t_{0}^{2}$
is equivalent to $\left\Vert \left[2t_{0}\,(t_{1}-t_{2})\right]\right\Vert \le t_{1}+t_{2}$,
where we have $t_{1}\ge0$ and $t_{2}\ge0$. Based on the above preparations,
for the point ($\tilde{\mathbf{w}},\tilde{x}$), the problem (\ref{SRM3})
can then be transformed into a convex optimization problem as follows
\begin{subequations}\label{SRM3Tr}
\begin{align}
\max_{\{\mathbf{w}_{m}\}_{m=1}^{M},t_{0},t_{1},t_{2}} & \quad t_{0}\\
\mathrm{s.t.}\qquad & \quad1+\sum_{m=2}^{M}\mathbf{w}_{m}^{H}\mathbf{H}_{1}\mathbf{w}_{m}\nonumber\\&\quad\le Q_{\mathbf{H}_{1},1}(\mathbf{w}_{1},t_{1},\tilde{\mathbf{w}}_{1},\tilde{t}_{1}),\\
 & \quad1+\sum_{m=1}^{M}\mathbf{w}_{m}^{H}\mathbf{H}_{E}\mathbf{w}_{m}\nonumber\\&\quad\le\frac{2}{\tilde{t_{2}}}-\frac{t_{2}}{\tilde{t}_{2}^{2}}+\sum_{m=2}^{M}Q_{\mathbf{H}_{E},0}(\mathbf{w}_{m},t_{2},\tilde{\mathbf{w}}_{m},\tilde{t}_{2}),\\
 & \quad\gamma_{m}\left(1+\sum_{q=1,q\ne m}^{M}\mathbf{w}_{q}^{H}\mathbf{H}_{m}\mathbf{w}_{q}\right)\nonumber\\&\quad\le\mathbf{w}_{m}^{H}\mathbf{H}_{m}\mathbf{w}_{m},m\in\left[2,M\right],\\
 & \quad\sum_{m=1}^{M}\left\Vert \mathbf{w}_{m}\right\Vert ^{2}\le P_{M},\\
 & \quad\left\Vert \left[2t_{0}\,(t_{1}-t_{2})\right]\right\Vert \le t_{1}+t_{2}.
\end{align}
\end{subequations} Eventually, the problem can be converted to the following SOCP form \eqref{SOCP}
\begin{algorithm*}[t]
\begin{subequations}\label{SOCP}
\begin{align}
\max_{\{\mathbf{w}_{m}\}_{m=1}^{M},t_{0},t_{1},t_{2}}&\quad t_{0}\\
\mathrm{s.t.}\qquad& \quad\left\Vert \left[2\mathbf{w}_{2}^{H}\mathbf{h}_{1},\dots,2\mathbf{w}_{M}^{H}\mathbf{h}_{1},g_{1}-1\right]^{T}\right\Vert \le g_{1}+1,\\
& \quad\left\Vert \left[2\mathbf{w}_{1}^{H}\mathbf{h}_{E},\dots,2\mathbf{w}_{M}^{H}\mathbf{h}_{E},g_{2}-1\right]^{T}\right\Vert\le g_{2}+1,\\
& \quad\left\Vert \left[2\mathbf{w}_{1}^{H}\mathbf{h}_{m},\ldots,2\mathbf{w}_{m-1}^{H}\mathbf{h}_{m},2\mathbf{w}_{m+1}^{H}\mathbf{h}_{m},..,2\mathbf{w}_{M}^{H}\mathbf{h}_{m},2,g_{3m}-1\right]^{T}\right\Vert \le g_{3m}+1,\\
& \quad\mathrm{Im}(\mathbf{w}_{m}^{H}\mathbf{h}_{m})=0,m\in\left[2,M\right],\\
& \quad\left\Vert \left[\mathbf{w}_{1}^{T},\ldots,\mathbf{w}_{M}^{T}\right]^{T}\right\Vert \le\sqrt{P_{M}},\\
& \quad\left\Vert \left[2t_{0},(t_{1}-t_{2})\right]\right\Vert \le t_{1}+t_{2},
\end{align}
\end{subequations} %
\end{algorithm*}
where the linear functions $g_{1},g_{2},g_{3m}$ are respectively
defined as
\begin{equation}
g_{1}=Q_{\mathbf{H}_{1},1}(\mathbf{w}_{1},t_{1},\tilde{\mathbf{w}}_{1},\tilde{t}_{1})-1,\label{GG1}
\end{equation}
\begin{equation}
g_{2}=\sum_{m=2}^{M}Q_{\mathbf{H}_{E},0}(\mathbf{w}_{m},t_{2},\tilde{\mathbf{w}}_{m},\tilde{t}_{2})-t_{2}/\tilde{t}_{2}^{2}-1,\label{GG2}
\end{equation}
\begin{equation}
g_{3m}=\mathrm{Re}\left(\mathbf{w}_{m}^{H}\mathbf{h}_{m}\right)/\sqrt{\gamma_{m}},m\in\left[2,M\right].\label{GG3}
\end{equation}
It is noted the SOCP (\ref{SOCP}) can be solved efficiently by using the
available solvers such as $\mathtt{CVX}$ \cite{grant2010cvx}. The
proposed STB-OM scheme is summarized by Algorithm 1 as follows.
\vspace{-0.1cm}
\begin{algorithm}[tbh]
\begin{enumerate}
\item Initialization: Set $\mathbf{\tilde{w}}_{m}$, $\tilde{t}_{1}$ and
$\tilde{t}_{2}$ as the values which are feasible to the problem (\ref{SOCP}).
\item Repeat:\\
Solve the SOCP problem (\ref{SOCP}) with $(\mathbf{\tilde{w}}_{m},\tilde{t}_{1},\tilde{t}_{2})$
and obtain the optimal values $(\mathbf{w}_{m}^{*},t_{1}^{*},t_{2}^{*})$.\\
 Update $(\mathbf{\tilde{w}}_{m},\tilde{t}_{1},\tilde{t}_{2})=(\mathbf{w}_{m}^{*},t_{1}^{*},t_{2}^{*})$.
\item Until the convergence threshold is satisfied or the maximize number
of iterations is reached.
\item Output $\mathbf{w}_{m}$.
\end{enumerate}
\caption{STB-OM}
\end{algorithm}

The original optimization problem of STB-OM is nonconvex and NP-hard, thus its global optimum cannot be obtained with polynomial computational complexity using any known algorithm. By contrast, the proposed scheme, which is described by Algorithm 1, employs the first order Taylor approximations to transform the original optimization problem into a convex optimization problem. The globally optimal solution of the resultant convex optimization problem can be obtained upon using Algorithm 1, which, according to \cite{Vishwanathan05kernelmethods}, can be proved to converge to a local optimum of the original optimization problem in a few steps\footnote{The performance gap between Algorithm 1 and the optimal scheme cannot be provided at this stage due to the difficulty of obtaining the global optimum of the original optimization problem. Nevertheless, when proper initial values are given (although this is challenging as well), the local optimum achieved by Algorithm 1 may be equal to the global optimum of the original problem.}.

\begin{figure}[tbh]
\centering{}\includegraphics[width=8cm]{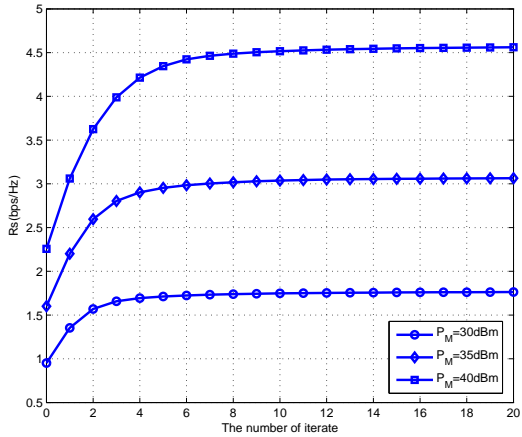}\protect\caption{\label{fig:Convergence-property-according}The convergence property of Algorithm 1 under different values of the transmit power of the MBS.}

\end{figure}

\begin{rem}
Since the optimization problem defined in (\ref{SOCP}) is convex,
the optimal solutions $\{\mathbf{w}_{m}^{*}\}_{m=1}^{M},t_{0}^{*},t_{1}^{*},t_{2}^{*}$,
are obtained by solving (\ref{SOCP}) for a given $(\mathbf{\tilde{w}}_{m},\tilde{t}_{1},\tilde{t}_{2})$.
At each iteration, $(\mathbf{\tilde{w}}_{m},\tilde{t}_{1},\tilde{t}_{2})$
is updated based on the optimal solution $(\mathbf{w}_{m}^{*},t_{1}^{*},t_{2}^{*})$
obtained in the previous iteration. Hence, $(\mathbf{\tilde{w}}_{m},\tilde{t}_{1},\tilde{t}_{2})$
is always a feasible solution of the current iteration, and the value of $t_{0}$ obtained for the given $(\mathbf{\tilde{w}}_{m},\tilde{t}_{1},\tilde{t}_{2})$
will be larger than or equal to the value of $t_{0}$ of the previous iteration.
This observation reveals that the required secrecy rate will be monotonically increasing (or at least nondecreasing) at each iteration, which is demonstrated by the numerical results in Fig. \ref{fig:Convergence-property-according}.
Due to the power constraint, there is an upper bound for the achievable secrecy rate. Therefore, the convergence of the proposed Algorithm 1 can be guaranteed.
\end{rem}

\begin{rem}
It is clear that for the STB-OM, only the local CSI is required at the MBS, including  the CSI of MUs and the CSI of the eavesdropper.
\end{rem}

\begin{rem}
If an FU in a femtocell is wiretapped by another FU in the same femtocell,
the problem is mathematically identical to that we have solved in
this subsection. In other words, the proposed approach is applicable
to general single-cell interference-free scenarios.
\end{rem}

\subsection{Secrecy Transmit Beamforming Sequentially Performed in Macrocell and
Femtocell (STB-SMF) \label{sub:Joint-Secrecy-Transmit}}

\begin{figure}[t]
\begin{centering}
\includegraphics[width=8cm]{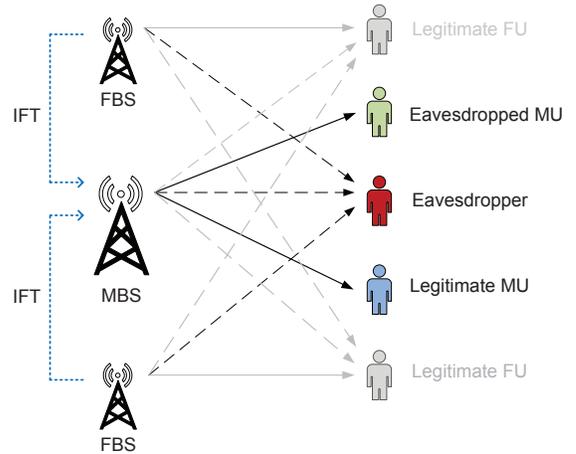}\caption{An example channel model for STB-SMF scheme, consisting of one MBS,
$M=2$ MUs, $N=2$ cooperative FBSs, and $K=1$ FU at each femtocell. An eavesdropper
wiretaps the first MU in the macrocell. As the SONOSA is adopted, the channel model is equivalent
to an interference channel model. The black solid lines indicate
useful data streams, and the black dash lines indicate interference
streams, which are optimized by STB-SMF. In particular, the blue dotted lines indicate the feedback of
interference temperatures from FBSs to the MBS. }

\par\end{centering}

\label{system_oneFBS}
\end{figure}

Compared to the conventional transmit beamforming without the consideration of
secrecy, the secrecy rate performance of $\text{MU}_{1}$
can be improved by the STB-OM for its clear objective. In fact, we may further improve
the secrecy rate performance by designing more sophisticated approaches. As shown in \cite{sheikholeslami_physical_2012}, it is potentially
feasible to exploit interference for improving the secrecy rate performance.
In order to achieve this goal, judicious interference management schemes could be developed. To justify these assumptions, a secrecy-oriented spectrum allocation strategy, i.e., the proposed SONOSA, is invoked to introduce deliberate CCI that is
friendly with respect to the secrecy rate performance of $\text{MU}_{1}$. The SONOSA strategy dynamically changes the local pattern of the underlay OSA to improve the performance of STB-OM.
Based on SONOSA, the cooperative co-channel FBSs can act independently as the sources of friendly CCI to degrade
the performance of the eavesdropper. Specifically, the cooperative FBSs selflessly maximize the power of leakage
interference imposed on the eavesdropper to improve the secrecy rate performance of $\text{MU}_{1}$. By exploiting the null space beamforming technique at the
cooperative FBSs, the interference would only affect the eavesdropper's received
signal.

Here, as shown in Fig. 4, we develop a novel STB-SMF scheme to exploit the interference in HetNet with SONOSA. In what follows, we
first discuss the STB design for the $n$-th cooperative FBS $\mathrm{\mathrm{FB}S}_{n}$, $n\in\left[1,N\right]$. We consider that $\mathrm{FBS}_{n}$ transmits data streams in the null space
of $\mathbf{G}_{n}=\left[\mathbf{h}_{n,1,}^{T}\mathbf{h}_{n,2}^{T},...,\mathbf{h}_{n,M}^{T}\right]^{T}\in\mathbb{C}^{M\times N_{F}}$, which is the collective channels from $\mathrm{\mathrm{FB}S}_{n}$ to all legitimate MUs.
To guarantee the existence of the null space,
$N_{M}>N_{F}>M$ has to be satisfied. As a consequence, the interference from $\mathrm{FBS}_{n}$ would
only degrade the eavesdropper's channel. In order to improve the secrecy rate, we aim at maximizing
the interference from $\mathrm{\mathrm{FB}S}_{n}$ to the eavesdropper subject to the
transmit power constraint. The problem can be
then formulated as follows
\begin{subequations}\label{eq:oo1}
\begin{align}
\max_{\mathbf{W}_{n}} & \quad \mathrm{Tr}\left[\mathbf{W}_{n}^{H}\mathbf{H}_{n,E}\mathbf{W}_{n}\right]\\
\mathrm{s.t.} &  \quad  \mathrm{Tr}\left[\mathbf{W}_{n}^{H}\mathbf{W}_{n}\right]\le P_{F}, \\
 & \quad  \mathbf{G}_{n}\mathbf{W}_{n}=\mathbf{0},
\end{align}
\end{subequations}
where $\mathbf{H}_{n,E}=\mathbf{h}_{n,E}^{H}\mathbf{h}_{n,E}$, $\mathbf{W}_{n}=\left[\mathbf{w}_{n,1},\mathbf{w}_{n,2},\dots,\mathbf{w}_{n,K}\right]$,
and $\mathbf{G}_{n}\mathbf{W}_{n}=\mathbf{0}$ ensures that $\mathrm{FBS}_{n}$
does not generate interference to the legitimate MUs. In order to eliminate the inter-cell inter-user interference of the $K$ $\mathrm{FU}$s served by $\mathrm{FBS}_{n}$, the beamforming vectors of $\mathrm{FBS}_{n}$
should also satisfy $\mathbf{w}_{n,k}^{H}\mathbf{h}_{n,nt}=0,\, k,t\in[1,K],\, k\ne t$.
Let us define $\mathbf{w}_{n,k}=\mathbf{V}_{n}\mathbf{x}_{n,k}$,
where the columns of $\mathbf{V}_{n}$ constitute an orthogonal basis
for the null space of $\mathbf{G}_{n}$. The optimization problem
(\ref{eq:oo1}) is then equivalent to
\begin{subequations}\label{Nullspace}
\begin{align}
\max_{\{\mathbf{x}_{n,k}\}_{k=1}^{K}} &  \quad \sum_{k=1}^{K}\mathbf{x}_{n,k}^{H}\mathbf{R}_{1n}\mathbf{x}_{n,k}\\
\mathrm{s.t.}\quad &   \quad \sum_{k=1}^{K}\mathbf{x}_{n,k}^{H}\mathbf{R}_{2n}\mathbf{x}_{n,k}\le P_{F}, \\
 &   \quad \mathbf{x}_{n,k}^{H}\mathbf{V}_{n}^{H}\mathbf{h}_{n,nt}=0, k,t\in[1,K], \,k\ne t,
\end{align}
\end{subequations}
where we have $\mathbf{R}_{1n}=\mathbf{V}_{n}^{H}\mathbf{H}_{n,E}\mathbf{V}_{n},$
and $\mathbf{R}_{2n}=\mathbf{V}_{n}^{H}\mathbf{V}_{n}.$ The above problem
can be further transformed into a SOCP problem by introducing a slack
variable $\alpha$ as follows
\begin{subequations}\label{eq:333}
\begin{align}
\max_{\{\mathbf{x}_{n,k}\}_{k=1}^{K},\alpha} & \quad\alpha\\
\mathrm{s.t.}\quad & \quad\left\Vert \left[\mathbf{h}_{n,E}\mathbf{V}_{n}\mathbf{x}_{n,1},\dots,\mathbf{h}_{n,E}\mathbf{V}_{n}\mathbf{x}_{n,K}\right]^{T}\right\Vert \le\alpha,\\
 & \quad\left\Vert \left[\mathbf{V}_{n}\mathbf{x}_{n,1},\dots,\mathbf{V}_{n}\mathbf{x}_{n,K}\right]^{T}\right\Vert \le\sqrt{P_{F}},\\
 & \quad\mathbf{x}_{n,k}^{H}\mathbf{V}_{n}^{H}\mathbf{h}_{n,nt}=0,\, k\ne t\,,\, k,\, t\in[1,K].
\end{align}
\end{subequations}
As we can see, the problem (\ref{eq:333}) is a standard SOCP problem
and can be solved efficiently via some numerical solvers, such as
$\mathtt{CVX}$ \cite{grant2010cvx}. For some special case, we can even obtain the closed-form
solution. For example, when the number of FU served by $\mathrm{\mathrm{FB}S}_{n}$ is 1, i.e., $K=1,$ the
problem (\ref{Nullspace}) can be simplified as
\begin{eqnarray}
\max_{\mathbf{x}_{n,1}} &  & \mathbf{x}_{n,1}^{H}\mathbf{R}_{1n}\mathbf{x}_{n,1}\label{Nullspace-1}\\
\mathrm{s.t.} &  & \mathbf{x}_{n,1}^{H}\mathbf{R}_{2n}\mathbf{x}_{n,1}\le P_{F}.\nonumber
\end{eqnarray}
The optimal solution of (\ref{Nullspace-1}) is given by $\mathbf{x}_{n,1}^{*}=\alpha\phi_{max}(\mathbf{R}_{1n},\mathbf{R}_{2n})$,
where $\phi_{max}(\mathbf{R}_{1n},\mathbf{R}_{2n})$ denotes the generalized
eigenvector corresponding to the largest generalized eigenvalue of
the matrix pair $(\mathbf{R}_{1n},\mathbf{R}_{2n})$ and $\alpha=\sqrt{P_{F}}\left(\left\Vert \mathbf{R}_{2n}^{\frac{1}{2}}\phi_{max}\left(\mathbf{R}_{1n},\mathbf{R}_{2n}\right)\right\Vert \right)^{-1}.$
The optimal value of the objective function (\ref{Nullspace-1}) is
$P_{F}\lambda_{max}(\mathbf{R}_{1n},\mathbf{R}_{2n})$, where $\lambda_{max}(\mathbf{R}_{1n},\mathbf{R}_{2n})$
is the largest generalized eigenvalue of matrix pair $(\mathbf{R}_{1n},\mathbf{R}_{2n})$.
As a result, the optimal beamforming vector can be expressed as $\mathbf{w}_{n,1}^{*}=\mathbf{V}_{n}\mathbf{x}_{n,1}^{*}.$

After designing the STB at the cooperative FBSs, we continue to design the STB at the MBS. It is noted that based on the interference temperature
generated by the cooperative FBSs (cf. (\ref{eq:SINRe})), we can obtain the STB of the MBS by employing the STB-OM with minor modification. Specifically, let us define the interference temperature generated by $\mathrm{FBS}_{n}$ at the eavesdropper as
\begin{equation}
\mathrm{IFT}_{n}=\sum_{k=1}^{K}\left|\mathbf{h}_{n,E}^{H}\mathbf{w}_{n,k}^{*}\right|^{2}, \label{IFT}
\end{equation}
then the sum of such interference from all $N$ cooperative FBSs is $\mathrm{IFT}_{sum}=\sum_{n=1}^{N}\mathrm{IFT}_{n}.$
It should be noted that $\mathrm{IFT}_{sum}$ is calculated by the MBS, after receiving the feedback about $\mathrm{IFT}_{n}$ from $\mathrm{FBS}_{n}$. Then the MBS performs STB according to STB-OM and $\mathrm{IFT}_{sum}$.
For clarity, the proposed STB-SMF is summarized by Algorithm 2 as follows.
\begin{algorithm}[tbh]
\begin{enumerate}
\item Obtain the beamforming vector $\mathbf{w}_{n,k}^{*}$ of $\mathrm{\mathrm{FB}S}_{n}$ with (\ref{eq:333}) or (\ref{Nullspace-1}).
\item Each cooperative $\mathrm{FBS}_{n}$ calculates $\mathrm{IFT}_{n}$ (\ref{IFT}) and transmit it to the MBS.
\item The MBS calculates $\mathrm{IFT}_{sum}=\sum_{n=1}^{N}\mathrm{IFT}_{n}$  and obtains the beamforming vector of the MBS according to the STB-OM in Algorithm
1.
\end{enumerate}
\caption{STB-SMF}
\end{algorithm}

\begin{rem}
It is worth mentioning that each cooperative FBS only needs to send a
scalar $\mathrm{IFT}_{sum}$ to the MBS in this scheme. Hence, STB-SMF
only incurs a small increase of the backhaul traffic loads. Furthermore, only the local CSI is required
at each FBS. In contrast, the local CSI and the interference temperature are required at the MBS.
\end{rem}

\begin{rem}
In the STB-SMF, we take the QoS constraints of MUs into consideration while the QoS
of FUs is ignored. In other words, we do not intend to ensure or enhance the QoS of FUs. Transmit beamforming vectors are designed for the
cooperative FBSs to improve the secrecy rate of the eavesdropped MU, and no interference is generated to the other MUs. The STB-SMF is an
altruistic manner from the view point of the FBSs.  A more practical scheme is developed in the next subsection, where the QoS of FUs is considered.
Despite this fact, the STB-SMF proposed in this subsection serves as a preliminary scheme which provides us the first insight into the interference-aided
physical layer security enhancement in HetNet. Furthermore, the STB-SMF scheme also acts as a design alternative for achieving the tradeoff between the performance and the implementation complexity. Compared to the practical scheme proposed in the next subsection, less cooperation and computation are required by the STB-SMF.
\end{rem}

\subsection{Secrecy Transmit Beamforming Jointly Performed in Macrocell and Femtocell (STB-JMF)}

\begin{figure}[t]
\begin{centering}
\includegraphics[width=8cm]{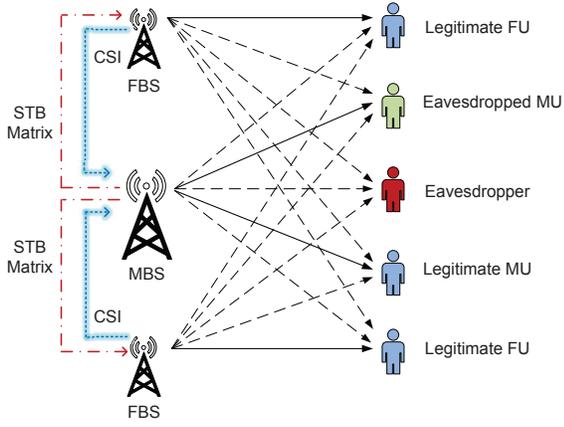}\caption{An example channel model for STB-JMF scheme, consisting of one MBS,
$M=2$ MUs, $N=2$ FBSs, and $K=1$ FU at each femtocell. An eavesdropper
wiretaps the first MU in the macrocell.  As the SONOSA strategy is used, the channel model is equivalent to an interference
channel model. The solid lines indicate useful
data streams, and the dash lines indicate the interference streams.
The blue glow dotted lines indicate the CSI streams delivered from
FBSs to the MBS. The red dash-dot lines denote the STB vectors delivered
from the MBS to the FBSs.}

\par\end{centering}

\label{system_twoFBS}
\end{figure}

In the STB-SMF scheme, CCI is deliberately introduced
and exploited. The secrecy rate performance of $\text{MU}_{1}$ is thus
enhanced, and the QoS of the other MUs is also guaranteed. Nevertheless,
the QoS of the co-channel FUs is ignored, which implies that we cannot
ensure the sum-rate performance of the FUs of the cooperative FBSs.

In this section, we propose a joint secrecy transmit beamforming scheme
named STB-JMF to improve the secrecy rate of $\text{MU}_{1}$ while
ensuring the QoS of the other MUs and FUs. Note that the SONOSA  is also used in the STB-JMF scheme
in order to take advantage of the CCI from the
point of secrecy. Our goal is to maximize the secrecy rate of $\text{MU}_{1}$ subject to the
MBS/FBSs transmit power constraints and the SINR requirements at the
other MUs and FUs. The transmit beamforming vectors of both the MBS
and the cooperative FBSs are optimized jointly. To be more specific, each cooperative FBS obtains its local
CSI and then sends it to the MBS. As a result, global CSI becomes available
at the MBS. Then, the MBS jointly optimizes the beamforming vectors with the aid of the global CSI.

Following the analysis in the previous sections, we can formulate
the original optimization problem as
\begin{subequations}\label{1}
\begin{align}
\max_{\begin{subarray}{c}
\{\mathbf{w}_{m}\}_{m=1}^{M},\\\{\{\mathbf{w}_{nk}\}_{k=1}^{K}\}_{n=1}^{N}\end{subarray}} & \quad\log_{2}(1+\mathrm{SINR}_{1})-\log_{2}(1+\mathrm{SINR}_{E})\label{eq:1a}\\
\mathrm{s.t.}\qquad\quad\;  & \quad\sum_{m=1}^{M}\|\mathbf{w}_{m}\|^{2}\le P_{M},\nonumber\\&\quad \sum_{k=1}^{K}\|\mathbf{w}_{nk}\|^{2}\le P_{F},n\in\left[1,N\right],\label{eq:1b}\\
 & \quad\mathrm{SINR}_{m}\ge\gamma_{m},m\in\left[2,M\right],\label{eq:1c}\\
 & \quad\mathrm{SINR}_{nk}\ge\gamma_{nk},n\in\left[1,N\right],k\in\left[1,K\right],\label{eq:1d}
\end{align}
\end{subequations}
where \eqref{eq:1b} characterizes the total transmit
power constraints at the MBS and the cooperative FBSs, \eqref{eq:1c} is the QoS requirement
of intact user $m$ in the macrocell, and \eqref{eq:1d} is the QoS
requirement of $\mathrm{FU}_{nk}$. As we can see, the optimization
problem is non-convex and hence is very hard to solve. Despite the
challenge, in what follows we will show that the problem can be solved
globally optimally by reformulating it as a two-part problem. Let \textbf{$\mathbf{W}_{m}=\mathbf{w}_{m}\mathbf{w}_{m}^{H},$
$\mathbf{W}_{nk}=\mathbf{w}_{nk}\mathbf{w}_{nk}^{H},$ } we can use
semidefinite relaxation technique \cite{shaoshi_Yang_2013} to simplify
the problem. First, we introduce a slack variable $\tau=\mathrm{SINR}_{E}$, and
(\ref{1}) can be equivalently transformed into \begin{subequations}\label{2}
\begin{align}
\max_{\begin{subarray}{c}
\{\mathbf{w}_{m}\}_{m=1}^{M},\\\{\{\mathbf{w}_{nk}\}_{k=1}^{K}\}_{n=1}^{N}\end{subarray}} & \quad\frac{1}{1+\tau}\left[1+\frac{\mathrm{Tr}(\mathbf{H}_{1}\mathbf{W}_{1})}{A_{1}^{'}}\right]\\
\mathrm{s.t.}\qquad\quad\;\; & \quad\sum_{m=1}^{M}\mathrm{Tr}(\mathbf{W}_{m})\le P_{M},\label{eq:2a}\\
 & \quad\sum_{k=1}^{K}\mathrm{Tr}(\mathbf{W}_{nk})\le P_{F},n\in\left[1,N\right],\label{eq:2b}\\
 & \quad\frac{\mathrm{Tr}(\mathbf{H}_{m}\mathbf{W}_{m})}{A_{m}^{'}}\ge\gamma_{m},m\in\left[2,M\right],\\
 & \quad\frac{\mathrm{Tr}(\mathbf{H}_{n,nk}\mathbf{W}_{nk})}{C_{nk}^{'}}\ge\gamma_{nk},\nonumber\\&\quad n\in\left[1,N\right],k\in\left[1,K\right],\\
 & \quad\frac{\mathrm{Tr}(\mathbf{H}_{E}\mathbf{W}_{1})}{B^{'}}\le\tau,\label{eq:2c}
\end{align}
\end{subequations}where $A_{1}^{'}$, $A_{m}^{'}$, $B^{'}$ and
$C_{nk}^{'}$ are respectively defined as
\begin{align}
 & A_{1}^{'}=\sum_{m\ge2}^{M}\mathrm{Tr}(\mathbf{H}_{m}\mathbf{W}_{m})+\sum_{n=1}^{N}\sum_{k=1}^{K}\mathrm{Tr}(\mathbf{H}_{n,1}\mathbf{W}_{nk})+1,\label{eq:ex1}\\
 & A_{m}^{'}=\sum_{q=1,q\not=m}^{M}\mathrm{Tr}(\mathbf{H}_{m}\mathbf{W}_{q})+\sum_{n=1}^{N}\sum_{k=1}^{K}\mathrm{Tr}(\mathbf{H}_{n,m}\mathbf{W}_{nk})+1,\label{eq:ex2}\\
 & B^{'}=\sum_{m\ge2}^{M}\mathrm{Tr}(\mathbf{H}_{E}\mathbf{W}_{m})+\sum_{n=1}^{N}\sum_{k=1}^{K}\mathrm{Tr}(\mathbf{H}_{n,E}\mathbf{W}_{nk})+1,\label{eq:ex3}\\
 & C_{nk}^{'}=\sum_{t=1,t\not=k}^{K}\mathrm{Tr}(\mathbf{H}_{n,nk}\mathbf{W}_{nt})+\sum_{p=1,p\not=n}^{N}\sum_{t=1}^{K}\mathrm{Tr}(\mathbf{H}_{p,nk}\mathbf{W}_{pt})\nonumber \\
 &\qquad +\sum_{m=1}^{M}\mathrm{Tr}(\mathbf{H}_{nk}\mathbf{W}_{m})+1,\label{eq:ex4}
\end{align}
$\mathbf{H}_{n,m}=\mathbf{h}_{n,m}\mathbf{h}_{n,m}^{H},$ $\mathbf{H}_{n,E}=\mathbf{h}_{n,E}\mathbf{h}_{n,E}^{H}$,
$\mathbf{H}_{n,nk}=\mathbf{h}_{n,nk}\mathbf{h}_{n,nk}^{H},$ $\mathbf{H}_{p,nk}=\mathbf{h}_{p,nk}\mathbf{h}_{p,nk}^{H}$
and $\mathbf{H}_{nk}=\mathbf{h}_{nk}\mathbf{h}_{nk}^{H}.$

To solve the problem (\ref{2}), we divided it into two parts. The
outer part is a one-dimensional line search problem with $\tau$,
i.e.,
\begin{eqnarray}
\max_{\tau} &  & \frac{1+G(\tau)}{1+\tau}\label{eq:4}\\
\mathrm{s.t.} &  & 0\le\tau\le\mathrm{Tr}(\mathbf{H}_{1})P_{M}.\nonumber
\end{eqnarray}
The function $G(\tau)$ is defined by another optimization problem
to be described later. The lower bound about $\tau$ can be obtained
directly from \eqref{eq:2c}, while the upper bound is derived from
the fact that the secrecy rate is greater than or equal to
zero, hence $\tau\le\mathrm{Tr}(\mathbf{H}_{1})P_{M}.$ For a fixed
$\tau$, the inner part can be expressed as
\begin{eqnarray}
\max_{\{\mathbf{W}_{m}\}_{m=1}^{M},\{\{\mathbf{W}_{nk}\}_{k=1}^{K}\}_{n=1}^{N}} &  & \frac{\mathrm{Tr}(\mathbf{H}_{1}\mathbf{W}_{1})}{A_{1}^{'}}\label{eq:3}\\
\mathrm{s.t.} &  & \mathrm{\eqref{eq:2a}-\eqref{eq:2c},}\nonumber
\end{eqnarray}
and $G(\tau)$ is equal to the optimal value of (\ref{eq:3}).

It is observed that the problem (\ref{eq:4}) is equivalent to the
original problem (\ref{2}). For any fixed $\tau$, we can obtain
$G(\tau)$ by solving (\ref{eq:3}). Then, applying the one-dimensional
line search method, e.g. Golden Section Search, to the interval $[1,1+\mathrm{Tr}(\mathbf{H}_{1})P_{M}]$,
we can solve the problem (\ref{2}). Hence, the key step is to
solve (\ref{eq:3}) for a fixed $\tau$. In what follows, we will
concentrate on it.

Since, the objective function of (\ref{eq:3}) is a linear fractional
function and thus it is quasi-convex \cite{Cheung_2013}, then we
can use Charnes-Cooper transformation \cite{Cheung_2014} to convert
it into a linear one. Upon introducing the auxiliary variables $\mathbf{X}_{m},\mathbf{X}_{nk}\succeq0$,
$\zeta>0$, we can rewrite $\mathbf{W}_{m}$ and $\mathbf{W}_{nk}$
as $\mathbf{W}_{m}=\frac{\mathbf{X}_{m}}{\zeta}$, $\mathbf{W}_{nk}=\frac{\mathbf{X}_{nk}}{\zeta},\mbox{ }$ then the
problem (\ref{eq:3}) can be transformed into the problem (\ref{eq:5}).

\begin{algorithm*}[t]
\begin{subequations}\label{eq:5}
\begin{align}
\max_{\begin{subarray}{c}
\{\mathbf{X}_{m}\}_{m=1}^{M},\zeta\\
\{\{\mathbf{X}_{nk}\}_{k=1}^{K}\}_{n=1}^{N}
\end{subarray}} & \quad\mathrm{Tr}(\mathbf{H}_{1}\mathbf{X}_{1})\label{eq:50}\\
\mathrm{s.t.}\qquad & \quad\mathrm{\sum_{\mathit{m}=1}^{\mathit{M}}Tr}(\mathbf{X}_{m})\le P_{M}\zeta,\label{eq:5.1}\\
 & \quad\sum_{k=1}^{K}\mathrm{Tr}(\mathbf{X}_{nk})\le P_{F}\zeta,\label{eq:5.2}\\
 & \quad\mathrm{Tr}(\mathbf{H}_{m}\mathbf{X}_{m})\ge\gamma_{m}\big(\sum_{\begin{subarray}{c}
q=1,\\
q\not=m
\end{subarray}}^{M}\quad\mathrm{Tr}(\mathbf{H}_{m}\mathbf{X}_{q})+\sum_{n=1}^{N}\sum_{k=1}^{K}\mathrm{Tr}(\mathbf{H}_{n,m}\mathbf{X}_{nk})+\zeta\big),\label{eq: 5.3}\\
 & \quad\mathrm{Tr}(\mathbf{H}_{n,nk}\mathbf{X}_{nk})\ge\gamma_{nk}\big(\sum_{\begin{subarray}{c}
t=1,\\
t\not=k
\end{subarray}}^{K}\quad\mathrm{Tr}(\mathbf{H}_{n,nk}\mathbf{X}_{nt})+\sum_{\begin{subarray}{c}
p=1\\
p\not=n
\end{subarray}}^{N}\quad\sum_{t=1}^{K}\mathrm{Tr}(\mathbf{H}_{p,nk}\mathbf{X}_{pt})+\sum_{m=1}^{M}\mathrm{Tr}(\mathbf{H}_{nk}\mathbf{X}_{m})+\zeta\big),\label{eq:7} \\
 & \quad\mathrm{Tr}(\mathbf{H}_{E}\mathbf{X}_{1})\le\tau\big(\sum_{m=2}^{M}\mathrm{Tr}(\mathbf{H}_{E}\mathbf{X}_{m})+\sum_{n=1}^{N}\sum_{k=1}^{K}\mathrm{Tr}(\mathbf{H}_{n,E}\mathbf{X}_{nk})+\zeta\big),\label{eq:8}\\
 & \quad\sum_{m=2}^{M}\mathrm{Tr}(\mathbf{H}_{m}\mathbf{X}_{m})+\sum_{n=1}^{N}\sum_{k=1}^{K}\mathrm{Tr}(\mathbf{H}_{n,m}\mathbf{X}_{nk})+\zeta=1,\\
 & \quad\mathbf{X}_{m},\mathbf{X}_{nk}\succeq0,\zeta>0.\label{eq:9}
\end{align}
\end{subequations}
\end{algorithm*} It can be seen that the problem (\ref{eq:5}) is an SDP problem \cite{shaoshi_Yang_2013}, which can be solved efficiently
by using numerical solvers such as $\mathtt{CVX}$ \cite{grant2010cvx}%
. The optimal solution of the problem (\ref{eq:5}) is denoted by
$(\mathbf{X}_{m}^{*},\mathbf{X}_{nk}^{*},\zeta^{*})$. Hence the corresponding
optimal solution of the problem (\ref{eq:3}) can be obtained as $\mathbf{W}_{m}^{*}=\frac{\mathbf{X}_{m}^{*}}{\zeta^{*}}$,
\textbf{$\mathbf{W}_{nk}^{*}=\frac{\mathbf{X}_{nk}^{*}}{\zeta^{*}}$}.
Then, we can solve the problem (\ref{eq:4}) through the one-dimensional
line search method such as Golden Section Search. Note that to get
the finial solution, we need to solve a sequence of SDPs.

Let us denote the optimal solution of the problem (\ref{eq:4}) as
($\tilde{\mathbf{W}}_{m},\tilde{\mathbf{W}}_{nk})$. Then, we can
obtain the beamforming vector solution as follows: if $\mathrm{rank}(\tilde{\mathbf{W}}_{m})=1$,
the optimal beamforming vector $\tilde{\mathbf{w}}_{m}$ is exactly
obtained via eigenvalue decomposition; otherwise some rank-one approximation
procedures, e.g. Gaussian randomization \cite{luo2010semidefinite}
%
can be applied to $\tilde{\mathbf{W}}_{m}$ for obtaining $\tilde{\mathbf{w}}_{m}$.
The same procedure is applicable to $\mathbf{\tilde{W}}_{nk}$. For
the sake of clarity, the proposed STB-JMF is summarized in Algorithm 3.

\noindent\emph{Proposition} 1. The optimal solution $\{\mathbf{X}_{m}\}_{m=1}^{M}$, $\{\{\mathbf{X}_{nk}\}_{k=1}^{K}\}_{n=1}^{N}$
of Problem (\ref{eq:5}) is always rank-one.

\begin{IEEEproof}
Please see Appendix.
\end{IEEEproof}

Proposition 1 shows that using semidefinite relaxation is always tight and yields a rank-one solution for the STB-JMF.
\begin{algorithm}[t]
\begin{enumerate}
\item Initialization: Set $P_{M}$, $P_{F}$, $\gamma_{m}$, $\gamma_{nk}$,
$\varepsilon$ (the search precision of the Golden Section Search);
\item Compute $\tau_{max}=\mathrm{Tr}(\mathbf{H}_{1})P_{M}$;
\item Solve (\ref{eq:4}) by applying one-dimensional line search method,
e.g., the Golden Section Search, on interval $[1,\tau_{max}]$, obtaining
the optimal $\tau^{*}$. To do this, we have to solve (\ref{eq:5})
to obtain $G^{*}(\tau)$ for a fixed $\tau$.
\item Obtain $(\mathbf{X}_{m}^{*},\mathbf{X}_{nk}^{*},\zeta^{*})$ for $\tau^{*}$.
\item Let $\mathbf{W}_{m}^{*}=\frac{\mathbf{X}_{m}^{*}}{\zeta^{*}}$, $\mathbf{W}_{nk}^{*}=\frac{\mathbf{X}_{nk}^{*}}{\zeta^{*}}$.
\item If $\mathrm{rank}(\mathbf{W}_{m}^{*})=\mathrm{rank}(\mathbf{W}_{nk}^{*})=1$, we can obtain $\mathbf{w}_{m}^{*}$ and $\mathbf{w}_{nk}^{*}$ via eigenvalue
decomposition;
\item Otherwise we apply Gaussian randomization method to $\mathbf{W}_{m}^{*}$ and $\mathbf{W}_{nk}^{*}$
for finding approximate $\mathbf{w}_{m}^{*}$ and $\mathbf{w}_{nk}^{*}$.
\item End
\end{enumerate}
\caption{STB-JMF}
\end{algorithm}

\begin{rem}
In the STB-JMF, not only the QoS of MUs but also the QoS of FUs are
taken into consideration. In other words, we can guarantee the QoS
of all the legitimate users in both the macrocell and the cooperative femtocells. This is pragmatically
attractive because QoS is one of the most important performance metrics
for practical HetNet.
\end{rem}
\begin{rem}
The STB-JMF scheme is capable of satisfying the QoS requirements as
well as enhancing the secrecy rate performance of the eavesdropped
MU, which is in contrast to the sole beamforming in STB-OM and the
sequential beamforming in STB-SMF. The cooperative
FBSs need to share their local CSI with the MBS. The CSI of all the MUs
and FUs are available at the MBS, so that the transmit beamforming
vectors for the MBS and FBSs are all designed at the MBS. Then, the
MBS delivers the related beamforming matrices to the cooperative FBSs.
\end{rem}
\begin{rem}
The proposed three STB schemes exploit the cooperation among network
nodes in varying degrees. In the STB-OM scheme, no cooperation
is used. In STB-SMF, each cooperative FBS designs transmit beamforming
vectors separately for the sake of enhancing the secrecy rate performance
of the eavesdropped MU. A scalar needs to be fed back to the MBS
from each cooperative FBS to assist the secrecy beamforming at
the MBS. In STB-JMF, all the cooperative FBSs have to share the CSI
with the MBS, and the transmit beamforming vectors for the MBS and
FBSs are jointly designed at the MBS to satisfy universal QoS requirements,
which cannot be supported by the STB-OM and STB-SMF schemes. Moreover, it is
worth pointing out that no data sharing is required for any of the proposed
STB schemes, which ensures low traffic load on the backhaul links.
\end{rem}

\begin{rem}
Based on \cite{a.ben2001}, we derive the computational complexity
of the proposed algorithms as follows. Firstly, the computational complexity of Algorithm 1 is
$T_{I}\cdot\mathcal{O}(N_{M}M^{3.5}+N_{M}^{3}M^{2.5})\cdot\log_{2}(\frac{1}{\epsilon})$. Since Algorithm 2 is based on Algorithm 1
and the only difference is that Algorithm 2 has to first calculate the interference caused by each FBS,
the computational complexity of Algorithm 2 is then characterized by
$\big(N\cdot\mathcal{O}(N_{F}^{3}K+N_{F}K^{2})+T_{I}\cdot\mathcal{O}(N_{M}M^{3.5}+N_{M}^{3}M^{2.5})\big)\cdot\log_{2}(\frac{1}{\epsilon})$. Finally,
Algorithm 3 is based on SDP and one-dimensional line search, and its computational complexity is
$T_{S}\cdot\mathcal{O}\big(NK(N_{M}^{3.5}+N_{F}^{3.5})+N^{2}K^{2}(N_{M}^{2.5}+N_{F}^{2.5})+N^{3}K^{3}(N_{M}^{0.5}+N_{F}^{0.5})\big)\cdot\log_{2}(\frac{1}{\epsilon})$,
where $\epsilon$ denotes the accuracy requirement, $T_{I}$ is the number of iterations required in Algorithm 1 and 2, and $T_{S}$ is the number of searches carried out in Algorithm 3. As we can see, the computational complexity of Algorithm 3 is higher than that of Algorithm 1 and 2.
\end{rem}

\begin{rem}
It is easy to extend the proposed schemes to the scenario
where there are multiple eavesdroppers. The only difference is that the resultant optimization
problem has to consider the SINR constraints of multiple eavesdroppers. Then, we can use the proposed algorithms to solve them. For the scenario where multiple legitimate MUs are targeted by the eavesdroppers, we may maximize either the secrecy sum-rate of the network or the minimum secrecy rate of the legitimate MUs, and the process of solving them is similar to the proposed algorithms.
\end{rem}

\section{Simulation Results and Discussions}

In this section, simulation results are provided to evaluate the secrecy rate
of $\mathrm{MU}_{1}$ for different transmit beamforming schemes conceived. We consider
a downlink HetNet with a central MBS serving a circular region $C$
and the area of $C$ is denoted as $\mathcal{C}$. We suppose the
radius of $C$ is 500m. The FBSs are spatially distributed according
to a Poisson point process $\varphi_{f}$ with intensity
$\lambda_{f}$ \cite{nguyen_opportunistic_2012}. Therefore, the average
number of FBSs within the cellular coverage is given by $N_{FBS}=\lambda_{f}\mathcal{C}$.
Femtocells are derived from the Voronoi tessellation and we can attain
the stochastic geometry model\cite{heath_modeling_2013} for the cellular systems
illustrated in Fig. 2. The simplified system models of the three proposed STB schemes have been illustrated by Fig. 3, Fig. 4 and Fig. 5, respectively. In
all simulations, the antenna configurations of the MBS and the FBS are $N_{M}=10$ and
$N_{F}=4$, respectively. The number of MUs is $M=2$, and the number of FUs in each FBS is $K=1$. We also assume all MUs, FUs and the eavesdropper are single-antenna
nodes due to practical constraint. For STB-SMF and STB-JMF schemes, we assume there are two cooperative FBSs
helping improve the secrecy rate\footnote{Our numerical results, which are not provided here due to page limitations, show that two cooperative FBSs are enough for achieving good secrecy performance. This is because when the number of cooperative FBSs increases, the interference imposed on the wiretapped MU also increases, hence the secrecy performance may not see any notable improvement.}, and each cooperative FBS serves one FU. According
to the ITU-R channel simulation specifications\cite{itu_r_recommendation},
we assume that the MBS has larger transmit power than the FBSs and
the specific numerical values are given in the following. Moreover,
the channel is modeled as a Rayleigh fading channel in all the
simulations.

\subsection{Benchmark Scheme}
The benchmark scheme is designed as follows. First, we aim to maximize
the rate of the wiretapped MU, i.e., $\mathrm{MU}_{1},$ without the
consideration of secrecy, and obtain the non-secrecy-oriented beamforming
vectors for the legitimate MUs and the FUs within the cooperative femtocells. Then, we derive the secrecy rate of $\mathrm{MU}_{1}$ with the non-secrecy-oriented beamforming vectors, which serves as our benchmark scheme. To be more specific, this optimization problem can be expressed as\begin{subequations}
\begin{align}
\max_{\begin{subarray}{c}
\{\mathbf{w}_{m}\}_{m=1}^{M},\\\{\{\mathbf{w}_{nk}\}_{k=1}^{K}\}_{n=1}^{N}\end{subarray}}& \log_{2}(1+\mathrm{SINR}_{1})\\
\mathrm{s.t.}\qquad\quad\; &  \sum_{m=1}^{M}\|\mathbf{w}_{m}\|\le P_{M},\nonumber\\&\sum_{k=1}^{K}\|\mathbf{w}_{nk}\|\le P_{F},n\in[1,N], \label{eq:555}\\
 &   \mathrm{SINR}_{m}\ge\gamma_{m},m\in\left[2,M\right],\label{eq:wwe22}\\
 &  \mathrm{SINR}_{nk}\ge\gamma_{nk},n\in[1,N],k\in[1,K],\label{eq:weqefw}
\end{align}
\end{subequations}where \eqref{eq:555} characterizes the transmit power
constraint at the MBS and FBSs, while \eqref{eq:wwe22} and  \eqref{eq:weqefw}
represent the QoS requirements of the legitimate MUs and the FUs, respectively. Note that this nonconvex problem can be transformed into an SDP problem, which is convex, by dropping a rank-one constraint that emerges in the transformation process and has the form of $\text {rank}({\bf X}) = 1$. For more details, please refer to the process of transforming the problem (\ref{eq:3}) into the problem (\ref{eq:5}). Upon finishing the optimization, we can obtain the secrecy rate of $\mathrm{MU}_{1}$. This scheme has been considered by \cite{mukherjee2009fixed}, \cite{fakoorian2011mimo}.

\subsection{Performance of the Proposed Schemes}
\begin{figure}[t]
\begin{centering}
\includegraphics[width=8cm]{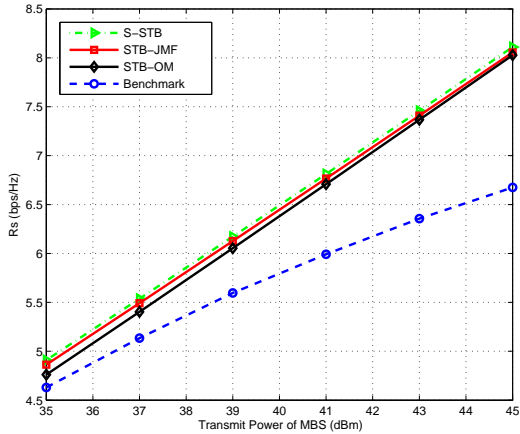}
\par\end{centering}

\caption{The secrecy rate of $\mathrm{MU}_{1}$ versus the transmit power of the
MBS. In the STB-SMF and STB-JMF schemes, two FBSs are used to generate interference.}

\label{sim_fig1}
\end{figure}

Fig. \ref{sim_fig1} shows the comparison results regarding  $\mathrm{MU}_{1}$'s secrecy rate performances
with different schemes for $P_{F}=40$dB.
Because the wiretapped $\mathrm{MU}_{1}$ receives  more power but the
received noise power does not rise when the transmit power of the MBS increases, so we can observe that the
secrecy rate performance of all the four schemes grow as the transmit power
of MBS increases.
It is also shown that the secrecy rates of the proposed three STB schemes always
increase faster than the benchmark scheme, especially when the transmit
power of the  MBS is high (e.g., 42dBm and 45dBm). Obviously,  the secrecy rate performance of
STB-JMF and STB-SMF is better than that of STB-OM. This observation
verifies that the deliberately introduced  interference is capable of enhancing  enhance the secrecy rate performance of the eavesdropped MU. Furthermore,
it is worth noting that $R_{s}$ of STB-SMF slightly outperforms STB-JMF
at all the transmit power values of the MBS, since STB-SMF does
not consider the QoS of FUs.

\begin{figure}[t]
\begin{centering}
\includegraphics[width=8cm]{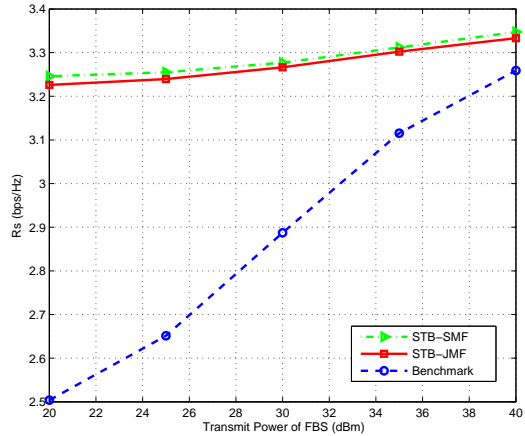}
\par\end{centering}

\caption{The secrecy rate of $\mathrm{MU}_{1}$ versus the transmit power of each
FBS.}

\label{sim_fig3-1}
\end{figure}

\begin{figure}[t]
\begin{centering}
\includegraphics[width=8cm]{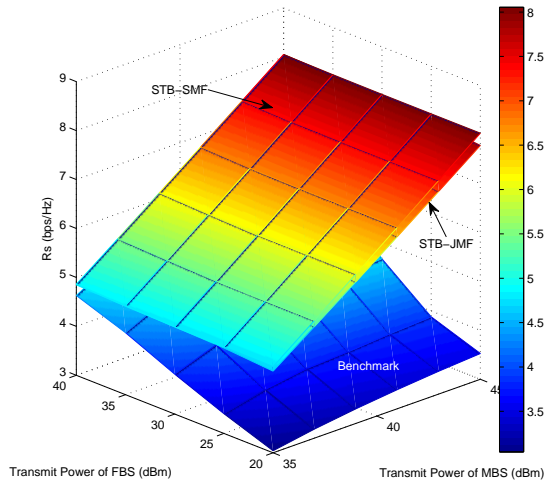}
\par\end{centering}

\caption{The secrecy rate region of $\mathrm{MU}_{1}$ versus the transmit
power of the MBS and each FBS.}

\label{sim_fig5}
\end{figure}

Fig. \ref{sim_fig3-1} illustrates the secrecy rate of $\mathrm{MU}_{1}$
for the STB-SMF and STB-JMF schemes under various FBS transmit power values.
It is shown that in all schemes the secrecy rate of $\mathrm{MU}_{1}$ improves as we increase the transmit power of FBS.
Note that the benchmark scheme could finally catch up our schemes at the cost of very high FBS power.
However, even with little transmit power at each FBS, the proposed schemes can achieve very high secrecy rate, which is in sharp contract to the benchmark scheme.
This indicates that STB-SMF and STB-JMF are
able to achieve a better $R_{s}$ without the need to increase the
transmit power of each cooperative FBS. In order to better illustrate the behaviors
of STB-SMF and STB-JMF, we plot the secrecy rate of $\mathrm{MU}_{1}$
versus the transmit power of the MBS and one cooperative FBS in a three-dimensional figure in Fig. \ref{sim_fig5}, where we can
observe the variation tendency of secrecy rate more clearly.

\begin{figure}[t]
\begin{centering}
\includegraphics[width=8cm]{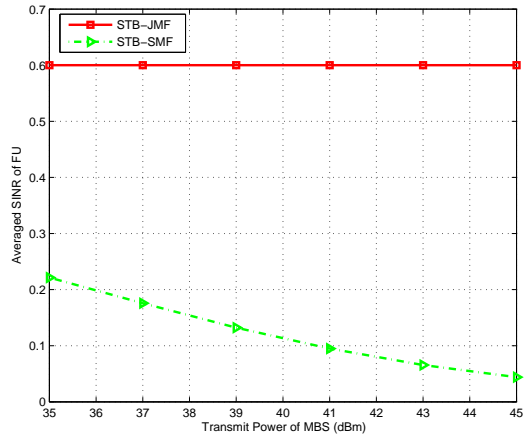}
\par\end{centering}

\caption{The receive SINR of FUs versus the transmit power of the MBS.}

\label{sim_fig6}
\end{figure}

Fig. \ref{sim_fig6} shows that the averaged
SINR of FUs versus the transmit power of the MBS when there is an eavesdropper
 wiretapping $\mathrm{MU}_{1}$.
Due to symmetry, we only need to evaluate the averaged SINR of the
FU in one cooperative femtocell, and we assume the SINR requirement of FUs in STB-JMF scheme
is 0.60. From Fig. \ref{sim_fig6}, it is observed that the SINR of
FUs in STB-JMF achieves its optimum and is significantly superior to that in STB-SMF. We can
also conclude that the SINR constraints \eqref{eq:1d} in STB-JMF
hold with equality.
But for the STB-SMF scheme, as the transmit power of the MBS increases, the SINR
of FUs goes down dramatically because more interference is introduced at FUs.

\begin{figure}[t]
\centering{}\includegraphics[width=8cm]{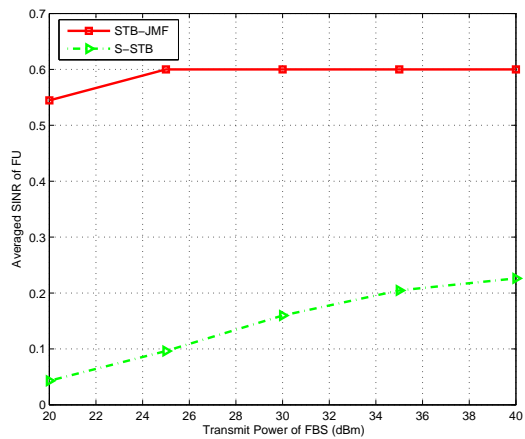}\caption{The receive SINR of FUs versus the transmit power of each FBS.}
\label{sim_fig7}
\end{figure}

Fig. \ref{sim_fig7} shows the averaged SINR of each FU versus the transmit
power of each FBS when there is an eavesdropper wiretapping $\mathrm{MU}_{1}$.
Similarly, the SINR requirement of FUs in STB-JMF is set as 0.60. We can
also see that STB-JMF has advantages related to the averaged SINR of
FU.
It is worth noting that when the transmit power of each FBS is relatively low, e.g., around 20 dBm,
the SINR constraint in STB-JMF could not be satisfied and thus there
is no optimal solution. Hence, we slightly decrease the SINR requirement
of FUs to 0.5 to ensure that the optimization problem can be solved. So in Fig.
\ref{sim_fig7}, at the point of 20dBm, we can see the averaged SINR
decreases slightly. However, from Fig. \ref{sim_fig6} and \ref{sim_fig7}, we can clearly see
that the benefit of STB-JMF is to maximize secrecy rate of the wiretapped
MU while maintaining QoS for all related FUs and the intended MU.

\subsection{Impact of Artificial Noise (AN) on the Proposed Schemes}
AN plays an important role in the physical layer security and may achieve substantial secrecy
performance in some scenarios. In what follows we provide some simulations to verify whether AN is still beneficial for our schemes. To this end, AN is introduced into the STB-OM, the STB-SMF and the STB-JMF in these simulations.
Furthermore, except for the beamforming schemes proposed in this paper, we also present
two additional schemes for the purpose of comparison in these simulations, which are the ``joint beamforming and AN design'', as well as the ``beamforming design with random AN''.

Let us consider the broadcast channel as an example, then the optimization problem of ``joint beamforming and AN design'' can be expressed as
\begin{subequations}\label{As-to-the}
\begin{align}
\max_{\{\mathbf{w}_{m}\}_{m=1}^{M},\mathbf{z}} & \log(1+\mathrm{SINR}_{1})-\log(1+\mathrm{SINR}_{E})\\
\mathrm{s.t.}\quad & \mathrm{SINR}_{m}\ge\gamma_{m},m\in\left[2,M\right],\\
 & \sum_{m=1}^{M}\left\Vert \mathbf{w}_{m}\right\Vert ^{2}+\left\Vert \mathbf{z}\right\Vert \le P_{M},
\end{align}
\end{subequations}
where $\mathrm{SINR}_{m}$ and $\mathrm{SINR}_{E}$ are the received SINRs at $\mathrm{MU}_{m}$ and at the
eavesdropper $E$, respectively. These SINRs are given by
\begin{equation}
\mathrm{SINR}_{m}=\frac{\left|\mathbb{\mathbf{h}}_{m}\mathbf{w}_{m}\right|^{2}}{\sigma_{M}^{2}+\sum_{q=1,q\neq m}^{M}\left|\mathbb{\mathbf{h}}_{m}\mathbf{w}_{q}\right|^{2}+\left|\mathbb{\mathbf{h}}_{m}\mathbf{z}\right|^{2}},\label{eq:sinrm}
\end{equation}
\begin{equation}
\mathrm{SINR}_{E}=\frac{\left|\mathbb{\mathbf{h}}_{E}\mathbf{w}_{1}\right|^{2}}{\sigma_{E}^{2}+\sum_{m=2}^{M}\left|\mathbb{\mathbf{h}}_{E}\mathbf{w}_{m}\right|^{2}+\left|\mathbb{\mathbf{h}}_{E}\mathbf{z}\right|^{2}},\label{eq:sinre}
\end{equation}
where $\mathbf{z}$ is the AN vector. Hence, we can solve this optimization
problem using Algorithm 1 proposed in this paper. Note that this model mimics the effect of AN in the STB-OM.

On the other hand, considering the ``beamforming design with random AN'' for the broadcast channel, we transmit random AN and the optimized beamforming vectors at the MBS. The optimization problem is similar to (\ref{As-to-the}) except that the optimization
variables are only $\{\mathbf{w}_{m}\}_{m=1}^{M}$, which do not include $\mathbf{z}$.
\begin{figure}[t]
\begin{centering}
\includegraphics[scale=0.5]{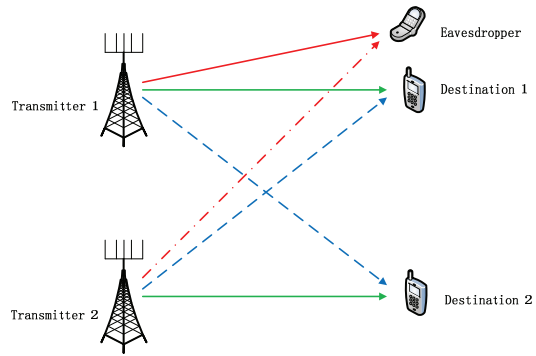}\protect\caption{\label{fig:system model}Two-user interference channel with an eavesdropper.}
\par\end{centering}

\end{figure}

\begin{figure}[t]
\begin{centering}
\includegraphics[width=8cm]{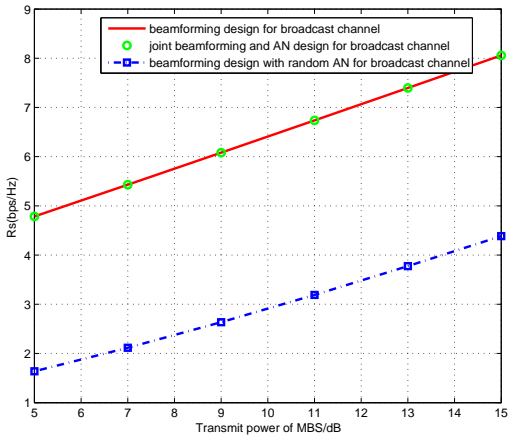}
\par\end{centering}
\centering{}\protect\caption{\label{Fig.1}The secrecy rate versus the transmit power of the MBS in the broadcast
channel with/without AN.}
\end{figure}

\begin{figure}[t]
\begin{centering}
\includegraphics[width=8cm]{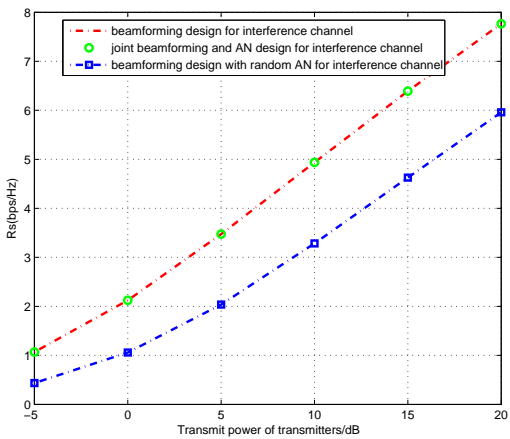}
\par\end{centering}
\centering{}\protect\caption{The secrecy rate versus the transmit power of transmitters in the interference channel
with/without AN.}
\label{Fig.2}
\end{figure}

Note that the STB-SMF and the STB-JMF rely on an interference channel. For simplicity, we only consider the typical scenario where there are two
transmitters each having four antennas, two receivers each having a single antenna
and one single-antenna eavesdropper trying to wiretap one of the transmitters,
as shown in Fig. \ref{fig:system model}.

We present the simulation results for the broadcast channel (as shown in Fig. \ref{Fig.1}) and for the interference channel (as shown in Fig.
\ref{Fig.2}). From Fig. \ref{Fig.1} and Fig. \ref{Fig.2}, we can see that the secrecy rate of the system is almost the same for the proposed schemes whether there exists AN (joint design) or not.
Therefore, we can conclude that transmitting AN using the additional $N_{M}-M$ dimensions does not have any significant impact on the performance of the proposed schemes. This is because
in the proposed STB-OM scheme, we assume that there are multiple MUs, thus the MBS has to transmit multiple data flows. Therefore, the STB-OM relies on a broadcast channel. At the eavesdropper, the data from other legitimate MUs can be regarded as interference. For the STB-SMF and the STB-JMF, the data from FBSs can also be treated as interference. These interferences are essentially equivalent to the special ANs imposed on the eavesdropper.

\section{Conclusions}

In this paper, we have investigated the physical-layer security schemes in a two-tier downlink HetNet. On the basis of the two suggested spectrum allocation strategies, the OSA and the SONOSA, three STB schemes, i.e., STB-OM, STB-SMF and STB-JMF, have been proposed to maximize the secrecy rate of the eavesdropped user. According to various considerations of the QoS requirements of the legitimate users, the three proposed secrecy schemes adopt different degrees of collaboration among MBS and its cooperative FBSs. In particular, the complicated nonconvex STB optimization problems have been solved by problem reformulations with SDP and SOCP techniques. Simulation results show the effectiveness of the proposed schemes. For the future work, it would be interesting to consider the scenario where multiple eavesdroppers and/or targeted MUs exist in the HetNet. Additionally, the robust STB schemes in the context of imperfect CSI may be investigated.

\appendix{}

\section*{Proof Of Proposition 1}

The maximization problem (\ref{eq:5}) can be equivalently transformed into
\begin{subequations}\label{sec:The-maximization-problem}
\begin{align}
\min_{\begin{subarray}{c}
\{\mathbf{X}_{m}\}_{m=1}^{M},\zeta\\
\{\{\mathbf{X}_{nk}\}_{k=1}^{K}\}_{n=1}^{N}
\end{subarray}} & \mathrm{Tr}(\mathbf{H}_{E}\mathbf{X}_{1})\label{eq:5-1}\\
\mathrm{s.t.}\qquad  & \mathrm{\sum_{\mathit{m}=1}^{\mathit{M}}Tr}(\mathbf{X}_{m})\le P_{M}\zeta,\label{eq:1-1}\\
 & \sum_{k=1}^{K}\mathrm{Tr}(\mathbf{X}_{nk})\le P_{F}\zeta,n\in[1,N],\label{eq:e}\\
 & \mathrm{Tr}(\mathbf{H}_{m}\mathbf{X}_{m})\ge\gamma_{m}\big(\sum_{\begin{subarray}{c}
q=1,\\
q\not=m
\end{subarray}}^{M}\mathrm{Tr}(\mathbf{H}_{m}\mathbf{X}_{q})\nonumber\\&+\sum_{n=1}^{N}\sum_{k=1}^{K}\mathrm{Tr}(\mathbf{H}_{n,m}\mathbf{X}_{nk})+\zeta\big),m\in[2,M],\\
 & \mathrm{Tr}(\mathbf{H}_{n,nk}\mathbf{X}_{nk})\ge\gamma_{nk}\big(\sum_{\begin{subarray}{c}
t=1,\\
t\not=k
\end{subarray}}^{K}\mathrm{Tr}(\mathbf{H}_{n,nk}\mathbf{X}_{nt})\nonumber\\&+\sum_{\begin{subarray}{c}
p=1\\
p\not=n
\end{subarray}}^{N}\sum_{t=1}^{K}\mathrm{Tr}(\mathbf{H}_{p,nk}\mathbf{X}_{pt})\nonumber \\
 & +\sum_{m=1}^{M}\mathrm{Tr}(\mathbf{H}_{nk}\mathbf{X}_{m})+\zeta\big),\nonumber\\&n\in[1,N],k\in[1,K],\\
 & \mathrm{Tr}(\mathbf{H}_{1}\mathbf{X}_{1})\ge\tau\big(\sum_{m=2}^{M}\mathrm{Tr}(\mathbf{H}_{m}\mathbf{X}_{m})\nonumber\\&+\sum_{n=1}^{N}\sum_{k=1}^{K}\mathrm{Tr}(\mathbf{H}_{n,1}\mathbf{X}_{nk})+\zeta\big),\\
 & \sum_{m=2}^{M}\mathrm{Tr}(\mathbf{H}_{E}\mathbf{X}_{m})\nonumber\\&+\sum_{n=1}^{N}\sum_{k=1}^{K}\mathrm{Tr}(\mathbf{H}_{n,E}\mathbf{X}_{nk})+\zeta=1,\label{eq:d}\\
 & \mathbf{X}_{m},\mathbf{X}_{nk}\succeq0,\zeta>0.
\end{align}
\end{subequations}
It is easy to verify that Problem (\ref{sec:The-maximization-problem})
satisfies the Slater's condition. Then, the Karush-Kuhn-Tucker (KKT)
conditions are the sufficient and necessary optimality conditions.
Some KKT conditions needed in the proof are expressed as follows:
\begin{subequations}
\begin{align}
 & \mathbf{G}_{1}^{*}=a^{*}\mathbf{I}+\mathbf{H}_{\mathrm{E}}+\sum_{m=2}^{M}b_{m}^{*}\gamma_{m}\mathbf{H}_{m}\nonumber\\&\qquad+\sum_{n=1}^{N}\sum_{k=1}^{K}c_{nk}^{*}\gamma_{nk}\mathbf{H}_{nk}-d^{*}\mathbf{H}_{1},\label{eq:KKT1}\\
 & \mathbf{G}_{1}^{*}\mathbf{X}_{1}=0,\label{eq:KKT2}\\
 & \mathbf{G}_{m}^{*}=a^{*}\mathbf{I}+\sum_{q\not=1,m}^{M}b_{q}^{*}\gamma_{q}\mathbf{H}_{q}+\sum_{n=1}^{N}\sum_{k=1}^{K}c_{nk}^{*}\gamma_{nk}\mathbf{H}_{nk}\nonumber\\&\qquad+d^{*}\tau\mathbf{H}_{m}+e^{*}\mathbf{H}_{\mathrm{E}}-b_{m}^{*}\mathbf{H}_{m},m\not=1,\label{eq:KKT3}\\
 & \mathbf{G}_{m}^{*}\mathbf{X}_{m}=0,m\not=1,\label{eq:KKT4}\\
 & \mathbf{G}_{nk}^{*}=a_{n}^{*}\mathbf{I}+\sum_{m=2}^{M}b_{m}^{*}\gamma_{m}\mathbf{H}_{n,m}+\sum_{t\not=k}^{K}c_{nt}^{*}\gamma_{nt}\mathbf{H}_{n,nt}\nonumber\\&\qquad+\sum_{p\not=n}^{N}\sum_{k=1}^{K}c_{pk}^{*}\gamma_{pk}\mathbf{H}_{n,pk}+d^{*}\tau\mathbf{H}_{n,1} -c_{nk}^{*}\mathbf{H}_{n,nk},\nonumber\\&\qquad n\in[1,N],k\in[1,K],\label{eq:KKT5}\\
 & \mathbf{G}_{nk}^{*}\mathbf{X}_{nk}=0,n\in[1,N],k\in[1,K],\label{eq:KKT6}
\end{align}
\end{subequations}where $\mathbf{G}_{m}^{*}\succeq\mathbf{0},$ $m\in[1,M]$,
$\mathbf{G}_{nk}^{*}\succeq\mathbf{0},$ $n\in[1,N],k\in[1,K]$, $a^{*}\ge 0,$
$a_{n}^{*}\ge 0,$ $b_{m}\ge 0,$ $c_{nk}\ge 0,$ $d\ge 0,$ $e\ge 0$ are the optimal dual
variables associated with the constraints $\mathbf{X}_{m},\mathbf{X}_{nk}\succeq\mathbf{0}$
and (\ref{eq:1-1})-(\ref{eq:d}), respectively.

Note that (\ref{eq:KKT1}), (\ref{eq:KKT3}) and (\ref{eq:KKT5}) have a similar structure, hence we only focus on the proof of $\mathrm{rank}(\mathbf{X}_{1}^{*})=1$.
$\mathrm{rank}(\mathbf{X}_{m}^{*})=1,$ $m\in[2,M],$ and $\mathrm{rank}(\mathbf{X}_{nk}^{*})=1,$
$n\in[1,N],k\in[1,K],$ can be proved by using the same method.

If $\mathbf{X}_{1}^{*}=\mathbf{0}$, the resultant secrecy rate is
zero, which is trivial. Then, we have $\mathbf{X}_{1}^{*}\neq0$.
According to \eqref{eq:KKT2}, the rank of $\mathbf{G}_{1}^{*}$
must be less than or equal to $n-1$, i.e.,
\begin{equation}
\mathrm{rank}(\mathbf{G}_{1}^{*})\le n-1.\label{eq:KKT7}
\end{equation}

An important observation is that with the optimal solution, the power
constraints (\ref{eq:1-1}) and (\ref{eq:e}) have to be satisfied
with equality \cite{boyd2004convex}. Therefore, we have $a^{*}>0$ and $a_{n}^{*}>0,n\in[1,N]$. Let
\begin{align}
\mathbf{V} & =a^{*}\mathbf{I}+\mathbf{H}_{\mathrm{E}}+\sum_{m=2}^{M}b_{m}^{*}\gamma_{m}\mathbf{H}_{m}+\sum_{n=1}^{N}\sum_{k=1}^{K}c_{nk}^{*}\gamma_{nk}\mathbf{H}_{nk}\nonumber \\
 & =\big(\mathbf{V}^{\frac{1}{2}}\big)^{2}\succ\mathbf{0},
\end{align}
where \textbf{$\mathbf{V}^{\frac{1}{2}}=\left(\mathbf{V}^{\frac{1}{2}}\right)^{H}\succ0,$
}then the rank of $\mathbf{G}_{1}^{*}$ can be further expressed as
\begin{align}
\mathrm{rank}(\mathbf{G}_{1}^{*}) & = \mathrm{rank}\left(\mathbf{V}-d^{*}\mathbf{H}_{1}\right)\nonumber \\
 & = \mathrm{rank}\left(\mathbf{V}^{\frac{1}{2}}\left(\mathbf{I}-d^{*}\mathbf{H}_{1}\right)\mathbf{V}^{\frac{1}{2}}\right)\nonumber \\
 & \overset{a}{=}  \mathrm{rank}\left(\mathbf{I}-d^{*}\mathbf{V}^{-\frac{1}{2}}\mathbf{H}_{1}\mathbf{V}^{-\frac{1}{2}}\right)\nonumber \\
 & \ge  n-1,\label{eq:KKT9}
\end{align}
where the equation $a$ holds true relying on the following fact
\begin{equation}
\mathrm{rank}(\mathbf{B})=\mathrm{rank}(\mathbf{AB})=\mathrm{rank}(\mathbf{BC})=\mathrm{rank}(\mathbf{ABC}),
\end{equation}
where $\mathbf{A}_{m\times m}$ and $\mathbf{C}_{n\times n}$ are both
non-singular matrices, and $\mathbf{B}_{m\times n}$ is an arbitrary
matrix.

From \eqref{eq:KKT7} and \eqref{eq:KKT9}, we have
$\mathrm{rank}(\mathbf{G}_{1}^{*})=n-1$. According to \eqref{eq:KKT2},
we have $\mathrm{rank}(\mathbf{X}_{1}^{*})\le\mathrm{dim}(\mathcal{N}(\mathbf{G}_{1}^{*}))=n-\mathrm{rank}(\mathbf{G}_{1}^{*})=1.$
Since $\mathbf{X}_{1}^{*}\neq\mathbf{0}$, we obtain $\mathrm{rank}(\mathbf{X}_{1}^{*})=1.$

Following the same procedure, we are capable of proving that $\mathrm{rank}(\mathbf{X}_{m}^{*})=1,$
$m\in[2,M],$ and $\mathrm{rank}(\mathbf{X}_{nk}^{*})=1,$ $n\in[1,N],k\in[1,K]$.
Hence, we have completed the proof.

\section*{Acknowledgement}

The authors would like to thank Jian Zhou, Haijing Liu and Deyue Zhang, who helped us improve the manuscript.

\bibliographystyle{IEEEtran}
\bibliography{ciationsv3,JSAC}

\begin{IEEEbiography}[{\includegraphics[width=1in,height=1.25in,clip,keepaspectratio]{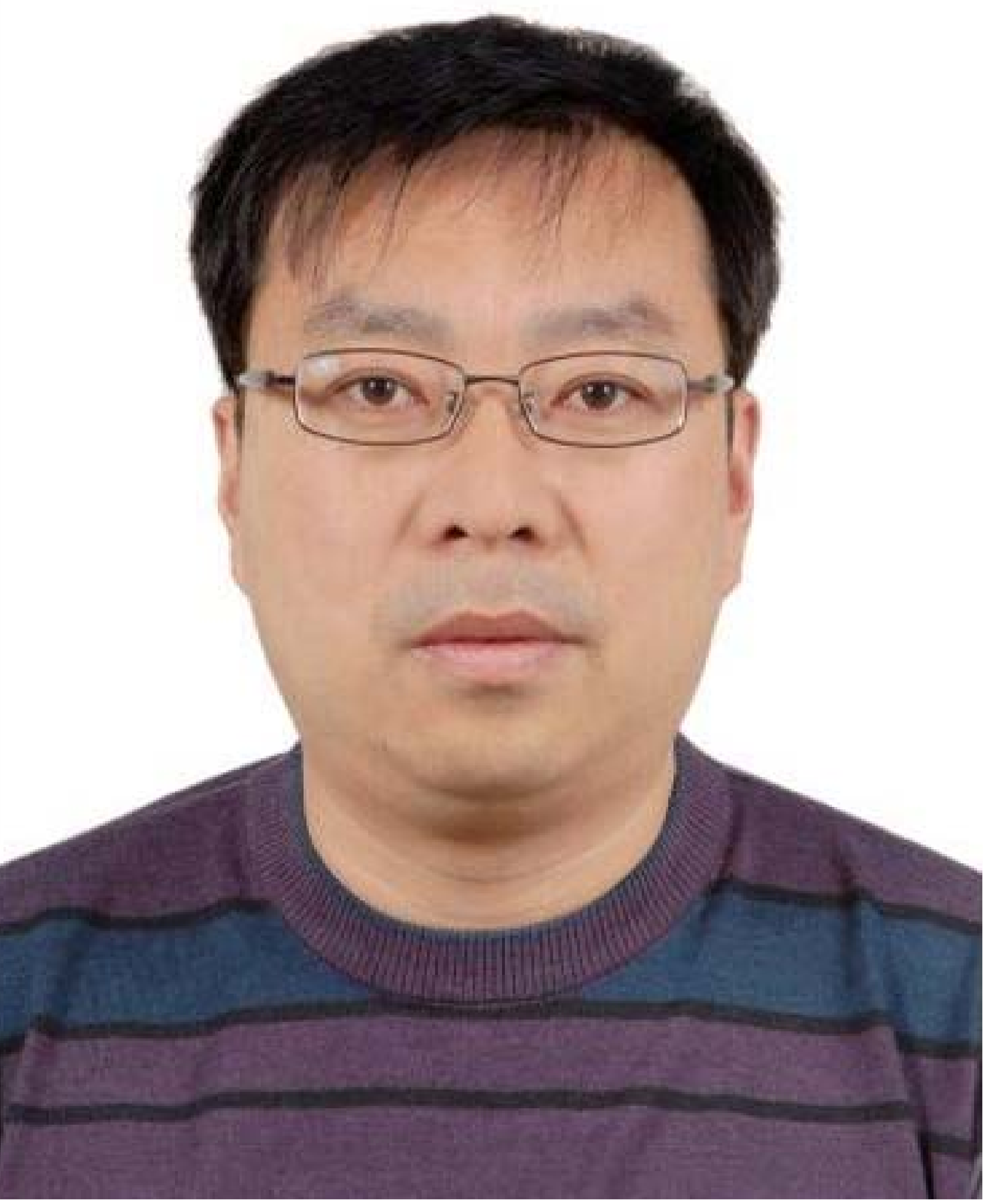}}]{Tiejun Lv}
(M'08-SM'12) received the M.S. and
Ph.D. degrees in electronic engineering from the
University of Electronic Science and Technology
of China (UESTC), Chengdu, China, in 1997 and
2000, respectively. From January 2001 to December
2002, he was a Postdoctoral Fellow with Tsinghua
University, Beijing, China. From September 2008
to March 2009, he was a Visiting Professor with
the Department of Electrical Engineering, Stanford
University, Stanford, CA. He is currently a Full
Professor with the School of Information and Communication
Engineering, Beijing University of Posts and Telecommunications
(BUPT). He is the author of more than 200 published technical papers on
the physical layer of wireless mobile communications. His current research
interests include signal processing, communications theory and networking.
Dr. Lv is also a Senior Member of the Chinese Electronics Association.
He was the recipient of the Program for New Century Excellent Talents in
University Award from the Ministry of Education, China, in 2006.
\end{IEEEbiography}

\begin{IEEEbiography}[{\includegraphics[width=1in,height=1.25in,clip,keepaspectratio]{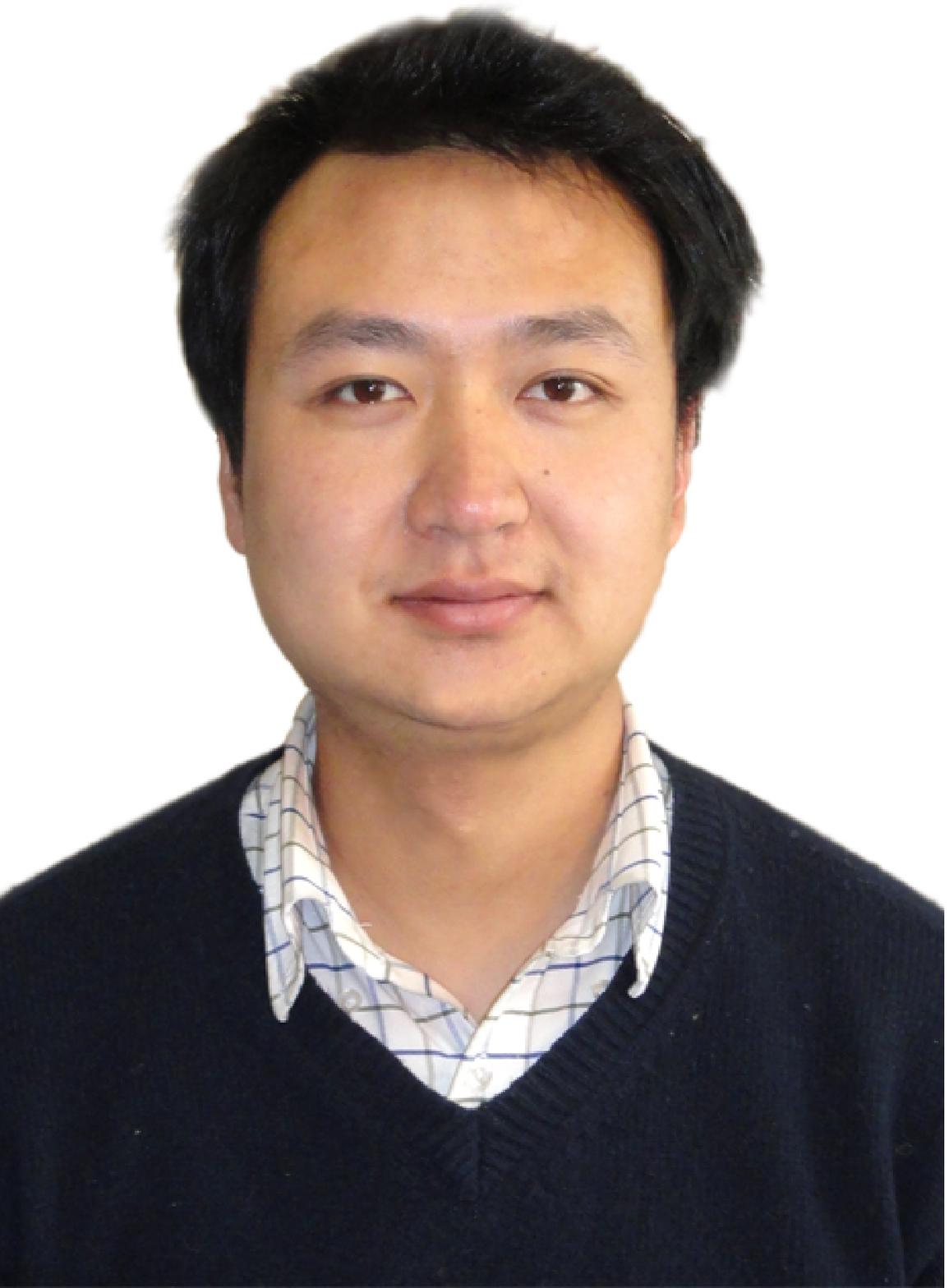}}]{Hui Gao}
(S'10-M'13) received his B. Eng. degree in Information Engineering and Ph.D. degree in Signal and Information Processing from Beijing University of Posts and Telecommunications (BUPT), Beijing, China, in July 2007 and July 2012, respectively. From May 2009 to June 2012, he also served as a research assistant for the Wireless and Mobile Communications Technology R\&D Center, Tsinghua University, Beijing, China. From Apr. 2012 to June 2012, he visited Singapore University of Technology and Design (SUTD), Singapore, as a research assistant. From July 2012 to Feb. 2014, he was a Postdoc Researcher with SUTD. He is now with the School of Information and Communication Engineering, Beijing University of Posts and Telecommunications (BUPT), as an assistant professor. His research interests include massive MIMO systems, cooperative communications, ultra-wideband wireless communications.
\end{IEEEbiography}

\begin{IEEEbiography}[{\includegraphics[width=1in,height=1.25in,clip,keepaspectratio]{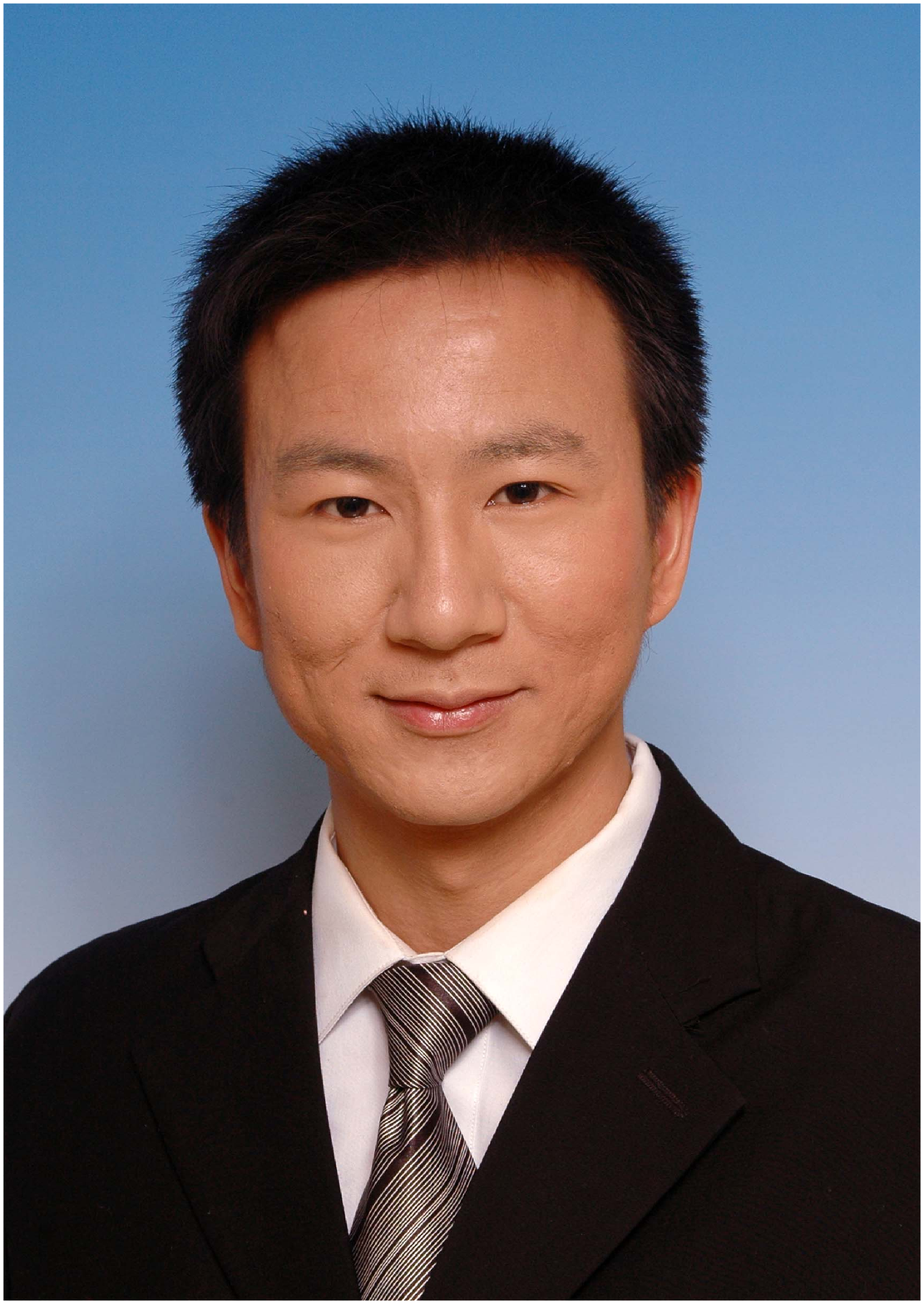}}]{Shaoshi Yang}
(S'09-M'13)  received
the B.Eng. Degree in Information Engineering
from Beijing University of Posts and
Telecommunications (BUPT), China, in 2006, the first Ph.D. Degree in Electronics and Electrical Engineering from University of Southampton, U.K., in 2013, and a second Ph.D. Degree in Signal and Information Processing from BUPT in 2014. Since 2013 he has been a Postdoctoral Research Fellow in
University of Southampton, U.K, and from 2008 to 2009, he was
an Intern Research Fellow with the Intel Labs China, Beijing, where he focused on
Channel Quality Indicator Channel design for mobile WiMAX (802.16 m).
His research interests include MIMO signal processing, green radio, heterogeneous networks, cross-layer interference management, convex optimization and its applications. He has published
in excess of 30 research papers on IEEE journals and conferences.

Shaoshi has received a number of academic and research awards, including the PMC-Sierra Telecommunications Technology Scholarship at BUPT, the Electronics and Computer Science (ECS) Scholarship of University of Southampton and the Best PhD Thesis Award of BUPT. He serves as a TPC member of a number of IEEE conferences and journals, including \textit{IEEE ICC, PIMRC, ICCVE, HPCC} and \textit{IEEE Journal on Selected Areas in Communications}. He is also a Junior Member of the Isaac Newton Institute for Mathematical Sciences, Cambridge University, UK. (https://
sites.google.com/site/shaoshiyang/)
\end{IEEEbiography}

\end{document}